\documentclass[preprint]{aastex}

\slugcomment{The Astrophysical Journal (in print).}
\usepackage{amsmath}

\shorttitle{Strong Gravity GR Tests with X-ray Polarimetry}
\shortauthors{Krawczynski}

\def\nerr#1#2 {{+{#1}}-{#2}}

\newcommand{\eq}[1]{Equation~(\ref{#1})}

\newcommand{\be}{\begin{equation}}
\newcommand{\ee}{\end{equation}}
\newcommand{\ba}{\begin{eqnarray}}
\newcommand{\ea}{\end{eqnarray}}

\begin{document}


\title{Tests of General Relativity in the Strong Gravity Regime Based on 
X-Ray Spectropolarimetric Observations of Black Holes in X-Ray Binaries}


\author{Henric Krawczynski}
\affil{Washington University in St. Louis,
Physics Department and McDonnell Center for the Space Sciences, 
1 Brookings Drive, CB 1105, St. Louis, MO 63130}


\begin{abstract}
Although General Relativity (GR) has been tested extensively in the weak gravity regime,
similar tests in the strong gravity regime are still missing. In this paper we explore the 
possibility to use X-ray spectropolarimetric observations of black holes
in X-ray binaries to distinguish between the Kerr metric and the phenomenological 
metrics introduced by Johannsen and Psaltis (2011) (which are not vacuum solutions 
of Einstein's equation) and thus to test the no-hair theorem of GR. 
To this end, we have developed a numerical code that calculates
the radial brightness profiles of accretion disks and parallel transports the wave vector
and polarization vector of photons through the Kerr and non-GR spacetimes. 
We used the code to predict the observational appearance of GR and non-GR accreting
black hole systems. We find that the predicted energy spectra and energy dependent polarization
degree and polarization direction do depend strongly on the underlying spacetime.
However, for large regions of the parameter space, the GR and non-GR metrics
lead to very similar observational signatures, making it difficult to observationally 
distinguish between the two types of models.
\end{abstract}
\section{Introduction}
The theory of General Relativity (GR) has been tested with high accuracy in the 
regime of weak gravity including tests within our solar system and tests based on observations 
of binary pulsars and the double pulsar \citep{Will:93,Will:06}.
Although X-ray energy spectra of mass accreting stellar mass and supermassive black holes
probe GR in the strong-gravity regime, the statistical and systematic uncertainties are sufficiently 
large that the data collected so far constrain the parameters describing the observed systems
\citep[i.e. the black hole spin, see][]{Mill:07,Ross:07,McCl:11,Kulk:11,Gou:11}, but do not yet permit to perform 
sensitive tests of GR \citep[however, see][]{Bamb:11,Bamb:12}.   
In the next few years, it may become possible to confront GR with new types of experimental data. 
In this paper, we discuss non-imaging spectropolarimetric X-ray observations of stellar mass black holes enabled
by a mission like GEMS (Gravity and Extreme Magnetism SMEX) \citep{Blac:10} or BEST 
(Black Hole Evolution and Space Time) \citep{Kraw:12}.  
Other electromagnetic observations with the potential to test strong gravity GR 
include pulsar timing observations \citep{Wex:99}, radio imaging of  supermassive 
black holes \citep[e.g.][]{Doel:09}, time resolved X-ray observations of the Fe K-$\alpha$ fluorescent line from
galactic and extragalactic black holes \citep[e.g.][]{Reyn:03,Guai:09}, and
the observation of stars orbiting the supermassive black hole at the center of the Milky Way \citep{Will:08,Meri:10}.
The continued improvement of the sensitivity of ground-based gravitational wave detectors should 
enable us to observe gravitational waves from compact object mergers before the end of this decade
\citep[e.g.][and references therein]{Hugh:10}. The ring down following a merger event contains information 
about dynamical GR in the strong gravity regime, but high signal to noise ratio observations are needed 
to measure the ring down with the required accuracy.

Observations of phenomena close to the event horizon of stellar mass black holes can probe 
gravity in more than five order of magnitude deeper potential wells, and for scalar curvatures 
more than twelve orders of magnitude larger than the best observations today 
\citep[see][for recent reviews of strong GR tests]{Psal:08,Psal:11}.
GR makes clear predictions of how astrophysical black holes look like: they are described
by the Kerr metric, a family of vacuum solutions that depends on two parameters, the mass 
$M$ of the black hole, and the angular momentum per unit mass $a$ with $0<a\le 1$ 
(in geometric units) corresponding to astrophysical black holes, while $a> 1$ gives 
rise to a naked singularity and is believed not to correspond to physical realizations. 
A series of papers in the late sixties and seventies established the "no-hair-theorem" stating 
that the Kerr (and Kerr-Newman) family of solutions are the only stationary 
axially symmetric  vacuum solutions of Einstein's equation \citep[see][and references therein]{Robi:09}. 
One way of testing strong gravity GR thus consists  in testing if the Kerr solution indeed 
describes astrophysical black holes.
In this paper, the possibility to use spectropolarimetric observations of the X-ray emission from 
black holes in X-ray binaries to verify GR in the strong gravity regime is discussed.
For moderate accretion rates, the disks are believed to be thin and to be described 
to good approximation by the Novikov-Thorne equations \citep{Novi:73}, 
the relativistic version of the Shakura-Sunyaev equations \citep{Shak:73}.
True disks are probably somewhat brighter close to the innermost stable circular orbit (ISCO) 
of the accreting matter owing to a non-vanishing  torque at the ISCO \citep{Nobl:11, Penn:12}. 
The radiation from a black hole in the "thermal state" is strongly dominated 
by the thermal emission from the inner accretion disk \citep{Remi:06,Abra:11} 
and presents a good opportunity to test the validity of accretion disk models and possibly also 
to test the underlying spacetime.

Quantitative tests of GR should give limits on parameters quantifying the deviation of reality from the predictions of GR.  
In the weak-gravity regime, a large body of experimental data has been used to constrain the deviations 
of the parameters of the parametrized post-Newtonian formalism from their values predicted by GR. 
In the absence of a general parametrization of the deviation from the Kerr spacetime, one has to turn to 
specific alternatives.  Black hole metrics following from alternative theories of gravity have been 
discussed in the literature. 
\citet{Alie:05} explored a metric describing charged spinning black holes
localized on a 3-brane in the Randall-Sundrum braneworld. 
Slowly and rapidly spinning black holes in Chern-Simon gravity were studied by \citet{Yune:09},
\citet{Konn:09}, and \citet{Klei:11}. 
\citet{Pani:11} considered slowly rotating 
black holes in theories where the Einstein-Hilbert action is supplemented by a scalar field coupling to the quadratic 
algebraic curvature invariants. 
In this paper, we follow the pragmatic approach to compare the predictions from the Kerr metric with
 the predictions from the axially symmetric metric described by  \citet{Joha:11} (JP11). 
 The metric includes the Kerr metric as a limiting case, and depends on parameters that describe the 
 deviation from the Kerr metric. 
It can describe static and rapidly spinning black holes 
and does not exhibit pathologies like timelike closed loops outside the event horizon.  
The objective of this paper is to see if X-ray polarimetry has sufficient diagnostic power 
to constrain the deviations from the Kerr metric.

A number of authors have studied the polarization properties of the X-ray emission from accreting black holes.
The X-rays from a flat-space Newtonian accretion disk are expected to be polarized owing to  
Thomson scattering  \citep[e.g.][]{Chan:60,Ange:69,Losk:82,Suny:85}. \citet{Star:77}, 
\citet{Conn:80a}, and \citet{Conn:80b} showed that
accounting for GR effects, the polarization degree and polarization direction of the emission from a thin accretion disk
show a complex energy dependence which might be used to estimate the parameters describing the system, e.g.\ the inclination of the system and the mass of the black hole. \citet{Dovc:08} evaluated the impact of various atmospheric 
optical depths on the observed polarization signal. \citet{Li:09} studied the same problem with a focus on 
how the polarization information can be used to break model degeneracies, i.\ e. to constrain the inclination 
of the inner accretion disk. \citet{Laor:90} and \citet{Matt:93} analyzed the  polarization of the UV/soft X-ray 
emission from AGNs. \citet{Pout:96ApJ} calculated the polarization of emission which Compton scatters in a 
hot corona above the accretion disk, and \citet{Pout:96MNRAS} and \citet{Dovc:04,Dovc:11} discussed the 
polarization of emission scattered by accretion disks. \citet{Dovc:06} and \citet{Zama:11} scrutinized the 
polarization signature  produced by orbiting hot-spots. \citet{Davi:09}, \citet{Sila:08} and \citet{Sila:11} studied the 
impact of turbulent magnetic fields on the observed polarization signatures. 
\citet{Schn:09} emphasized the importance of radiation scattered after returning to the disk 
owing to the spacetime curvature in the surrounding of the black hole on the observed polarization
\citep[see][for earlier related studies]{Cunn:76,Agol:00}. The same authors studied the observational 
appearance of the hard and steep power-law (SPL) states when coronal emission cannot be neglected \citep{Schn:10}. 

In this paper, we extend the work of \citet{Schn:10} to cover not only Kerr spacetimes 
but also the metric of JP11. 
Section \ref{S:Methodology} will discuss the technical aspects of our calculations, including the equations
used to derive the radial structure of the accretion disk, the formalism adopted to calculate the
polarization of the emission, and the ray tracing code. Section \ref{S:Results} will describe the results
of the simulations including the predicted observational signatures of the non-GR metrics.
We end with a discussion of the results in Sect.\ \ref{S:Discussion}.

We use the notation of \citet[][MTW73]{Misn:73}. Distances are given in units of the gravitational radius 
$r_{\rm g}\,=G\,M/c^2$, and we set $G=c=\hbar=1$ throughout the paper.
Einstein's summing convention is used. Greek indices run from 0 to 3 and Latin indices from 1 to 3.
\section{Methodology}
\label{S:Methodology}
\subsection{The Metric, Stable Circular Orbits, and Transformation Matrices} 
The following treatment assumes that  Einstein's Equivalence Principle (EEP) is valid. In brief: 
the trajectories of freely falling ``test'' bodies are independent of their internal structure and composition; 
the outcome of any local non-gravitational experiment is independent of the velocity of the freely-falling 
reference frame in which it is performed;  the outcome of any local non-gravitational experiment is independent 
of where and when in the Universe it is performed \citep[see][]{Will:06}.
We adopt the postulates of metric theories (the only theories which are consistent with the EEP) namely that
(i) spacetime is described by a symmetric metric, (ii) test bodies follow geodesics of this metric, and 
(iii) the non-gravitational laws of physics are consistent with the theory of special relativity in 
local freely falling reference frames. 

The metric of JP11 reads:
\[
ds^2 = -[1+h(r,\theta)] \left(1-\frac{2Mr}{\Sigma}\right)dt^2 -\frac{ 4aMr\sin^2\theta }{ \Sigma }[1+h(r,\theta)]dtd\phi + \frac{ \Sigma[1+h(r,\theta)] }{ \Delta + a^2\sin^2\theta h(r,\theta) }dr^2
\]
\be
+\Sigma d\theta^2 
 + \left[ \sin^2\theta \left( r^2 + a^2 + \frac{ 2a^2 Mr\sin^2\theta }{\Sigma} \right) + h(r,\theta) \frac{a^2(\Sigma + 2Mr)\sin^4\theta }{\Sigma} \right] d\phi^2
\label{E:metric}
\ee
with 
$\Sigma \equiv r^2 + a^2 \cos^2\theta$
and
$\Delta \equiv r^2 - 2Mr + a^2$.
The constants $M$ and $a$ are the mass and the spin of the black hole as measured by an observer at infinity. 

The function
\be
h(r,\theta) = \epsilon_3 \frac{M^3 r}{\Sigma^2}
\ee
parametrizes the deviation from the Kerr metric to leading order,  as lower orders are tightly constrained 
by the requirement for asymptotic flatness, the absence of gravitational waves, 
and Lunar Laser Ranging results  (see JP11).  For $\epsilon_3=0$  the metric reduces 
to the Kerr metric in Boyer Lindquist coordinates. 
We call $x^{\mu}=(t,r,\theta,\phi)$ the Global Coordinates (GCs) in the following, 
and denote the basis vectors of the tangent vector space at a point $\cal{P}$ as
${\bf e}_{\mu}\equiv$ $\partial/\partial x^{\mu}$.
The largest root of the equation $g_{t\phi}^{\,\,\,2}-g_{tt}g_{\phi\phi}\,=\,0$ gives the location of the event horizon,
and we infer the location of the ISCO from the condition: 
$dE/dr=0$.

The Killing vectors ${\bf e}_t$ and ${\bf e}_{\phi}$ lead to the conserved energy $E\equiv -p_t$ and
angular momentum $L_z\equiv p_{\phi}$ at infinity.  
JP11 derived these two quantities for particles of mass $\mu$ orbiting the black hole on circular geodesic 
orbits at radial coordinate $r$. The equations read (without loss of generality, we set $\mu=1$):
\be
E(r)\, = \frac{1}{r^6}\sqrt{ \frac{P_1 + P_2}{P_3} },\, 
\label{E:en}
\ee
\ba
L_z(r)\, = && \pm \frac{1}{r^4 P_6} \left[ \sigma \sqrt{\frac{M(r^3+\epsilon_3 M^3) P_5 }{P3}}  
\mp 6a M (r^3 + \epsilon_3 M^3) \sqrt{\frac{P_1 + P_2}{P_3}} \right] .
\label{E:am}
\ea
with the functions $P_1,\,...P_6$ given in Appendix A of JP11. The upper and lower signs
refer to co-rotating and counter-rotating particles, respectively. 
Compared to JP11, we modified Equ.\ (\ref{E:am}) to account for the facts that $(P_1 + P_2), P_3,$ and $P_5$ 
can be negative and $P_3$ should thus not be pulled out of the square root expressions.
Over most of the allowed $r$-range, $\sigma=1$; for some models  $P_5$ has an extremum with 
$P_5=0$ at $r_{\sigma}$ close to $r_{\rm ISCO}$ and $\sigma=-1$ has to be chosen for 
$r\le r_{\sigma}$ to make $L_z$ smooth and differentiable.
Equations (\ref{E:en}) and (\ref{E:am}), together with the equations
\be
p_{\theta}\,=0\,{\,\,\rm and\,\,}    p^2\,=\,-1 
\label{E:mom}
\ee
determine the four covariant (as well as the four contravariant) components of the momentum 
of the orbiting particles in GC. We restrict the discussion in this paper to prograde accretion disks.

The simulations require to transform tangent vectors like the photon wave vector ${\bf k}$ 
from the GC system to the local inertial frame of the orbiting particles, called the Plasma Frame (PF) in the following.   
We define an orthonormal basis vector system (indices with a hat) to treat emission and scattering processes 
in the  PF with one basis vector ${\bf e}_{\hat{t}}$ chosen to be parallel to the four velocity of the orbiting 
particles \cite[e.g.][]{Beck:08,Shch:11}. 
The four basis vectors are given by the equations:
\ba
{\bf e}_{\hat{t}}&\equiv&{\bf p}=p^t{\bf e}_t +p^{\phi}{\bf e}_{\phi},\\ \nonumber
{\bf e}_{\hat{r}}&\equiv&{\bf e}_r/{\sqrt{g_{rr}}},\\ \nonumber
{\bf e}_{\hat{\theta}}&\equiv&{\bf e}_{\theta} / {\sqrt{g_{\theta\theta}}}, \\ \nonumber
{\bf e}_{\hat{\phi}}&\equiv&\alpha\,{\bf e}_t +\beta\,{\bf e}_{\phi},
\ea
which define
the transformation matrices $e^{\nu}_{\rm\,\,\,\,\mu}$ through the relation
${\bf e}_{\hat{\mu}} =\,e^{\nu}_{\,\,\,\,\hat{\mu}}\,{\bf e}_{\nu}$.
The two constants $\alpha$ and $\beta$ are the positive solutions of the 
equations ${\bf e}_{\hat{t}} \cdot {\bf e}_{\hat{\phi}}\,=0$ and  
 ${\bf e}_{\hat{\phi}} \cdot {\bf e}_{\hat{\phi}}\,=1$.
We denote the inverse transformation matrices with a bar:
${\bf e}_{\nu} =\,\bar{e}^{\hat{\mu}}_{\,\,\,\,\nu}\,{\bf e}_{\hat{\mu}}$.
The components of a photon's wave vector {\bf k} then transform from the GC to the PF according to 
\be
k^{\hat{\mu}}=\bar{e}^{\hat{\mu}}_{\,\,\,\,\nu}\,k^{\nu},
\ee
and from the PF to the
GC according to 
\be
k^{\nu}=e^{\nu}_{\,\,\,\,\hat{\mu}}\,k^{\hat{\mu}}.
\ee
\subsection{Radial Structure of the Thin Accretion Disk}
We consider standard thin disk models with zero torque at the ISCO. 
For this case \citet{Page:74} (called PT74 in the following) showed that mass, energy, and angular momentum 
conservation alone determine the radial brightness profile of the accretion disk - 
exactly as in the case of a Shakura-Sunyaev accretion disk. 
PT74 derive the results for the general case of an axially symmetric metric
given in the equatorial plane of the accretion disk in the form
\be
ds^2\,=\,-e^{2\nu} dt^2+e^{2\psi} (d\phi-\omega dt)^2+e^{2\mu} dr^2+dz^2.
\label{E:PT}
\ee
The conservations laws can then be used to show that  the time average flux of radiant energy
(energy per unit proper time and unit proper area) flowing out of the upper surface of the disk, 
as measured by an observer on the upper face who orbits with the time-average motion of the disk's matter, 
is given by the equation:
\be
F(r)\,=\,\frac{\dot{M}_0}{4\pi}e^{-(\nu+\psi+\mu)} f(r)
\label{E:PT2}
\ee
with $\dot{M}_0$ being the radius independent time averaged rate at which rest mass flows inward through the disk.
The function $f$ depends on the momentum {\bf p} of the orbiting particles:
\be
f(r)\,\equiv\,
\frac{-p^{t}_{\,\,,r}}{p_{\phi}}\,
\int_{r_{\rm ISCO}}^{r}\,
\frac{p_{\phi,r}}{p^{t}}\,dr
\label{E:PT3}
\ee
where ``,'' denotes ordinary partial differentiation and $r_{\rm ISCO}$ is the radius of the ISCO.   
Comparison of \eq{E:PT} with \eq{E:metric} for $\theta=90^{\circ}$ and with $z$ properly defined, 
gives the functions $\nu$, $\psi$, $\mu$, and $\omega$ as function of $r$.
As $p_{\mu}(r)$ and $p^{\mu}(r)$ can be inferred from 
Equations (\ref{E:en}-\ref{E:mom}),  it is straight forward to solve Equations (\ref{E:PT2}) and (\ref{E:PT3}) numerically.
We cross-checked the results by testing that the emitted luminosity integrated over the entire accretion disk 
equals $\dot{M}_0$ times the radiative efficiency $\left(1-E(r_{\rm ISCO})\right)$.

The disk is assumed to have a temperature of 
\be
T_{\rm eff}=\left(\frac{F}{\sigma_{\rm SB}}\right)^{1/4}
\ee
with $\sigma_{\rm SB}$ the Stefan Boltzman constant, and to emit a diluted blackbody 
spectrum with a hardening factor of $f_{\rm h}=1.8$ (see below).
\subsection{Photon Emission, Ray Tracing, and Scattering of the Polarized Photons}
Our code simulates the emission of photons from the accretion disk, 
the photon propagation through the spacetime, and the 
scattering of photons returning to the accretion disk. 
For each geodesic, the code keeps track of statistical 
information including for example the temperature of the disk segment which emitted the photon. 
This statistical information is used when the observed energy spectra are calculated.
Although the discussion below uses the term photon, our treatment corresponds to the simulation of 
statistical ensembles of photon wave packages. 
In the following, $x^{\mu}$ refers to the approximate location of a wave package,  and ${\bf k}$, $\Pi$, and {\bf f} 
denote the mean wave vector, polarization degree and polarization vector of the wave package, 
respectively (see MTW73, paragraph 22.5 and \citet{Gamm:12} for a discussion of the {\bf k} and {\bf f}).

The numerical code simulates $n$ photons emitted from the plane of the accretion disk ($\theta=90^{\circ}$)
in $m$ radial bins logarithmically spaced from $r_{\rm ISCO}$ to $r_{\rm max}=100$, with $n=500$ and  
$m\,=\,10,000$.  Owing to the azimuthal symmetry of the problem, 
it is sufficient to simulate photons originating at $\phi=0$. 
The photons are launched into the upper hemisphere ($\theta<90^{\circ}$) 
with constant probability per solid angle in the PF, and with an initial wave vector {\bf k}$_0$ 
normalized such that $k_0^{\,\,\hat{t}}=|k_0^{\,\,\hat{i}}|=1$ in the PF. 
We use Table XXIV of \citep[][called C60 in the following]{Chan:60} 
for the polarized emission from an optical thick atmosphere to calculate the statistical weight $w_{\rm em}$ 
for the chosen emission direction in the PF, and to calculate the initial polarization $\Pi$ of the photon.
The polarization vector $f^{\hat{\mu}}$ is initialized by setting $f^{\hat{t}}=0$ and 
choosing $f^{\hat{i}}$ normalized to one and perpendicular to $k^{\hat{i}}$ and $(${\bf e}$_{\hat{\theta}})^{\hat{i}}$.
After calculating the components of {\bf k} and {\bf f} in the PF,  they are transformed into the GC system.
The polarization degree is an invariant.

The wave vector {\bf k} and polarization vector {\bf f}  are parallel transported with a similar algorithm as 
described by \citet{Psal:12}, extended to transport not only {\bf k} but also {\bf f}.
The two Killing vectors $\xi_1\,=\,(1,0,0,0)$ and  $\xi_2\,=\,(0,0,0,1)$ 
of the stationary axially symmetric metric imply the conservation of  the photon energy and angular momentum  
at infinity: 
\ba
E_{\gamma} &\equiv&  -k_t  =-g_{tt} \frac{dt}{d\lambda} -g_{t\phi} \frac{d\phi}{d\lambda} \label{E:cc1} \\ 
L_{\gamma} & \equiv&  k_{\phi}=g_{\phi\phi}\frac{d\phi}{d\lambda}+g_{t\phi} \frac{dt}{d\lambda} \label{E:cc2}
\ea
with  the affine parameter $\lambda$. 
These two equations are combined with ${dt}/{d\lambda}=k^t$ and ${d\phi}/{d\lambda}=k^{\phi}$
to calculate  $E_{\gamma}$ and $L_{\gamma}$ at the starting point of a geodesic (and also after the 
photon has been scattered).
A fourth order Runge-Kutta algorithm is used to integrate the geodesic equation: 
\be
\frac{d^2 x^{\mu}}{d\lambda'^2}\,=\,
-\Gamma^{\mu}_{\,\,\,\sigma\nu}
\frac{dx^{\sigma}}{d\lambda'}\frac{dx^{\nu}}{d\lambda'},
\ee
and to parallel transport the polarization vector according to the equation:
\be
\frac{d f^{\mu}}{d\lambda'}\,=\,
-\Gamma^{\mu}_{\,\,\,\sigma\nu}
f^{\sigma}\frac{dx^{\nu}}{d\lambda'}
\ee
with $\Gamma^{\mu}_{\,\,\,\sigma\nu}$ denoting the Christoffel symbols  and
$\lambda'\equiv E_{\gamma}\lambda$.

Equations (\ref{E:cc1}-\ref{E:cc2}) lead to the equations
${dt}/{d\lambda'}=$ $(-g_{\phi\phi}-b g_{t\phi})/d$ and 
${d\phi}/{d\lambda'}=$ $(b g_{tt} +g_{t\phi})/d$ with 
$b\equiv L_{\gamma}/E_{\gamma}$ and $d\equiv g_{\phi\phi} g_{tt}-g_{t\phi}^{\,\,\,2}$
which can be used to calculate ${dt}/{d\lambda'}$ and ${d\phi}/{d\lambda'}$
without the need for performing the Runge-Kutta integration. 
Note that the complex-valued Walker-Penrose integral of motion \citep{Walk:70} that can be used 
to simplify the parallel transport of {\bf f} through the Kerr metric \citep{Conn:80a,Conn:80b} 
was derived for vacuum solutions of Einstein's equation and cannot be used 
for the the metric of Equ.\ \ref{E:metric}.  The accuracy of the integration is monitored with the
invariants $\Delta_1=k^2$,
$\Delta_2=f^2-1$, and $\Delta_3=\bf k \cdot f$ along the photon trajectory. 

We assume that photons can cross the equatorial plane between the event horizon and the 
ISCO without interacting with the matter plunging into the black hole. 
If a photon hits the accretion disk it is scattered using Chandrasekhar's
formalism for scattering by a indefinitely deep atmosphere (C60, Section 70.3).
After calculating the components of the wave and polarization vectors in the PF, 
the Stokes and Chandrasekhar parameters are computed. The direction of the scattered photon 
in the PF is drawn from a random distribution with equal probability per solid angle and 
Equation (164) and Table XXV from C60 are used to calculate 
the outgoing Chandrasekhar parameters. 
Referring to Fig.\ 8 of C60, a beam with Stokes parameters ($Q=0$, $U=1$) correspond 
to a 45$^{\circ}$ clockwise rotation of the polarization direction relative to the direction with ($Q=1$, $U=0$) 
when looking towards the origin of the coordinate system for both, the incoming and the outgoing 
beams.

As an indefinitely deep atmosphere backscatters all incoming radiation and does not absorb any energy, 
it is clear that for each incoming photon direction given by the polar coordinates $\theta_0$, $\phi_0$, 
the average statistical weight averaged over all outgoing photon directions $\theta$, $\phi$ should be unity. 
Chandrasekhar's formula gives the polarized components of the outgoing intensity $I$
(energy emitted per unit time, accretion disk area, and solid angle) as function of the polarized
components of the incoming energy flux $\pi F$ per unit area perpendicular to the direction of the incoming photon.
The expression of the statistical weight for a certain scattering process reads:
\be 
w_{\rm sc}\,=\,\frac{2\pi \mu I}{ \mu_0 \pi  F} \,=\,\frac{2 \mu I}{ \mu_0 F}.
\ee
The factor of $2\pi$ in the numerator of the second expression converts from ``probability per solid angle'' to ``probability'',
the factor $\mu$ converts $I$ into the outgoing energy flux per solid angle and accretion disk area, and
the factor $\mu_0$ in the denominator converts $\pi F$ into the incoming energy flux per accretion disk area.
We verified that averaging $w_{\rm sc}$ over all directions in the upper hemisphere indeed gives unity.  
The simulation of a scattering event is completed by normalizing the temporal and spatial components of 
$k^{\hat{\mu}}$ after scattering to the value before scattering (both in the PF), and calculating the
GC components of {\bf k} and {\bf f}.
\subsection{Analysis of the Simulated Events}
Photons are tracked until they reach a distance of $r_{\rm r}=10,000$ or until they get too close to 
the event horizon. Once a photon reaches $r=r_{\rm r}$, the components of the wave vector 
{\bf k$_{\rm r}$} and the Stokes parameters are computed in the reference frame of a receiving 
observer with fixed coordinates.  We present below most results in the coordinate frame of this 
observer with momentum ${\bf p_{\rm r}}={\bf e}_t/\sqrt{g_{tt}}$. 
The tangent vectors are given in terms of components (marked with a tilde) with regards to the orthonormal basis:
\ba
{\bf e}_{\tilde{t}}&\equiv& {\bf p_{\rm r}}\,=\,{\bf e}_t/\sqrt{g_{tt}},\\ \nonumber
{\bf e}_{\tilde{r}}&\equiv& {\bf e}_r/\sqrt{g_{rr}},\\ \nonumber
{\bf e}_{\tilde{\theta}}&\equiv& {\bf e}_r/\sqrt{g_{\theta\theta}}, 
\ea
and with ${\bf e}_{\tilde{\phi}}$ being a linear combination of ${\bf e}_t$ and ${\bf e}_{\phi}$ 
normalized to one and orthogonal to $ {\bf e}_{\tilde{t}}$.
Photons arriving in the lower hemisphere are mirrored into the upper hemisphere to compensate for the fact 
that the simulations cover only emission into the upper hemisphere. The process leaves the Stokes parameter $Q$ 
invariant and flips the sign of $U$. 

Photons are accumulated in four $\pm$4$^{\circ}$-wide theta bins. Each photon contributes to the 
results with a statistical weight of:
\be
w_{\rm st}\,=\, 
2\pi \Delta r \frac{dN}{dt\,dr\,d\phi}
w_{\rm em} \,W_{\rm sc}.
\label{E:w}
\ee
Here, $\Delta r$ is the width of the radial bin from which a photon was launched, and $dN/dr\,dt\,d\phi$ is the 
number of photons emitted per GC unit time, radius, and azimuthal angle at the position of the radial bin.
The factors $w_{\rm em}$ and $W_{\rm sc}$ are the statistical weights for the emission process and 
all scattering processes, respectively. The latter equals the product of the weights $w_{\rm sc}$ of all 
scattering events if the photon was scattered, and 1 if not.
We calculate $dN/dr\,dt\,d\phi$ using a similar calculation as the one described in Appendix B2 of \citep{Kulk:11}. 
The number of photons emitted per area $d\hat{A}$ and time $d\hat{t}$ as measured in the PF is: 
\be
\frac{dN}{d^3\hat{V}}\, = \,\frac{F}{<\!\!\hat{E}\!>}
\ee 
with $d^3\hat{V}\equiv d\hat{A}\,d\hat{t}$.
Assuming a diluted blackbody spectrum with a hardening factor of $f_{\rm h}=1.8$, the mean PF energy 
per photon is $<\!\!\hat{E}\!>\,\approx$  
$2.70\, f_{\rm h}\, k_{\rm B}\,T_{\rm eff}$.
From the facts that the proper four-volume is an invariant and that the proper distance along 
the $\theta$-direction is an invariant for boosts along the ${\bf e}_{\phi}$-direction, 
it follows that the proper 3-volume $\sqrt{-g_{tr\phi}}\,dt\,dr\,d\phi$ (with 
$g_{tr\phi}$ being the $t$-$r$-$\phi$ part of the metric) is also an invariant.
Taking into account that $dN$ is an invariant and that  $d^3\hat{V}\,=$ 
$\sqrt{-g_{tr\phi}}dt\,dr\,d\phi$, we get:
 \be
\frac{dN}{dt\,dr\,d\phi}\,=\,   \sqrt{-g_{tr\phi}} \frac{F}{<\hat{\!\!E\!}>}
\ee 
with $\sqrt{-g_{tr\phi}}\,=$ $r$ for the Kerr metric and $\sqrt{-g_{tr\phi}}\,=$ $(1+h(r))\,r$ for the JP11 metric. 

2-D maps of the emission are made using $k_{\rm r}^{\tilde{\theta}}/k_{\rm r}^{\tilde{t}}$ 
and $k_{\rm r}^{\tilde{\phi}}/k_{\rm r}^{\tilde{t}}$
to give the photon direction in the image plane.  Each simulated photon is used once, 
rotating a photon arriving at $\phi$ to the azimuthal direction of the observer 
$\phi_{\rm r}=0$.  No additional re-weighting is necessary 
for this approach, as one can see by considering that if a $\phi$-bin of a certain width $\Delta\phi$ was adopted 
to select observed events, each simulated photon could be used $N$ times with regularly spaced $\phi$-offsets from 
0 to $360^{\circ}$, choosing $N$ such that the  number of photons falling into the  $\phi$-bin equalled unity.

For all energy-resolved results, the photons contributes with a statistical weight of 
\be
w_{E_1,E_2}\,= \,w_{\rm st}\,\int_{E_1}^{E_2} \frac{E^2}{e^{E/\epsilon_0}-1} dE\, /
\int_{0}^{\infty} \frac{E^2}{e^{E/\epsilon_0}-1} dE
\ee
with $\epsilon_0=f_{\rm h}\, T_{\rm eff }\,k_{\rm r}^{\tilde{t}}/k_0^{\,\,\hat{t}} $ to a bin ranging from $E_1$ to $E_2$.
The factor $k_{\rm r}^{\tilde{t}}/k_0^{\,\,\hat{t}}$ gives the redshift (or blueshift) from Doppler shifts following 
the emission of the photon and from the scattering(s) of the photon, and the gravitational redshift.
We measure the polarization direction $\chi$  (i.e. the direction of the electric field vector)
from the projection of the spin axis of the black hole in the sky with
$\chi$ increasing for a clockwise rotation of the polarization direction when looking towards the black hole.  

The ray-tracing algorithm offers ample opportunity for consistency checks, i.e.\ 
$k^2=0$ and $f^2=1$ can be checked along the photon trajectory.
We compared the 2-D maps and energy spectra of the polarization degrees and polarization
direction from our code with those from Figs.\ 1-3 of \citep{Schn:09} and found excellent agreement.
Note that \citet{Schn:09} adopt a different definition of the polarization vector. 
They use $\chi=0$ for a polarization parallel to the disk and $\chi$ increases 
for a counter-clockwise direction of the polarization direction in the sky.
\section{Results}
\label{S:Results}
We simulated the 10 parameter combinations listed in Table \ref{T:metric} (Models A-J). We assume a 
10 $M_{\odot}$ black holes accreting at rates between 0.5 and 4 times $10^{18}$~g~s$^{-1}$. 
The accretion rates were chosen to give a disk luminosity $L_{\rm D}\,=\,\left[1-E(r_{\rm ISCO})\right] \dot{M}$ of 
10\% of the Eddington luminosity. We considered Kerr black holes with spins of 
$a=$ 0, 0.5, 0.9, and 0.99, and non-GR metrics with $a=0.5$ ($\epsilon_3=$ -30.6, -5, 2.5, and 6.3) 
and $a=0.99$ ($\epsilon_3=$ -5, -2.5). 
Model E ($a=0.5$, $\epsilon_3=\, -30.6$) was chosen to have an ISCO $r_{\rm ISCO}=10\,r_{\rm g}$ larger than that of a 
Schwarzschild black hole ($r_{\rm ISCO}=6\,r_{\rm g}$). Model H ($a=0.5$, $\epsilon_3$=6.3) was chosen to have exactly 
the same ISCO as the Kerr black hole with $a=0.5$ (both $r_{\rm ISCO}=2.32\,r_{\rm g}$).

\begin{figure}
\epsscale{.80}
\begin{minipage}{5cm}
\resizebox{8cm}{!}{\plotone{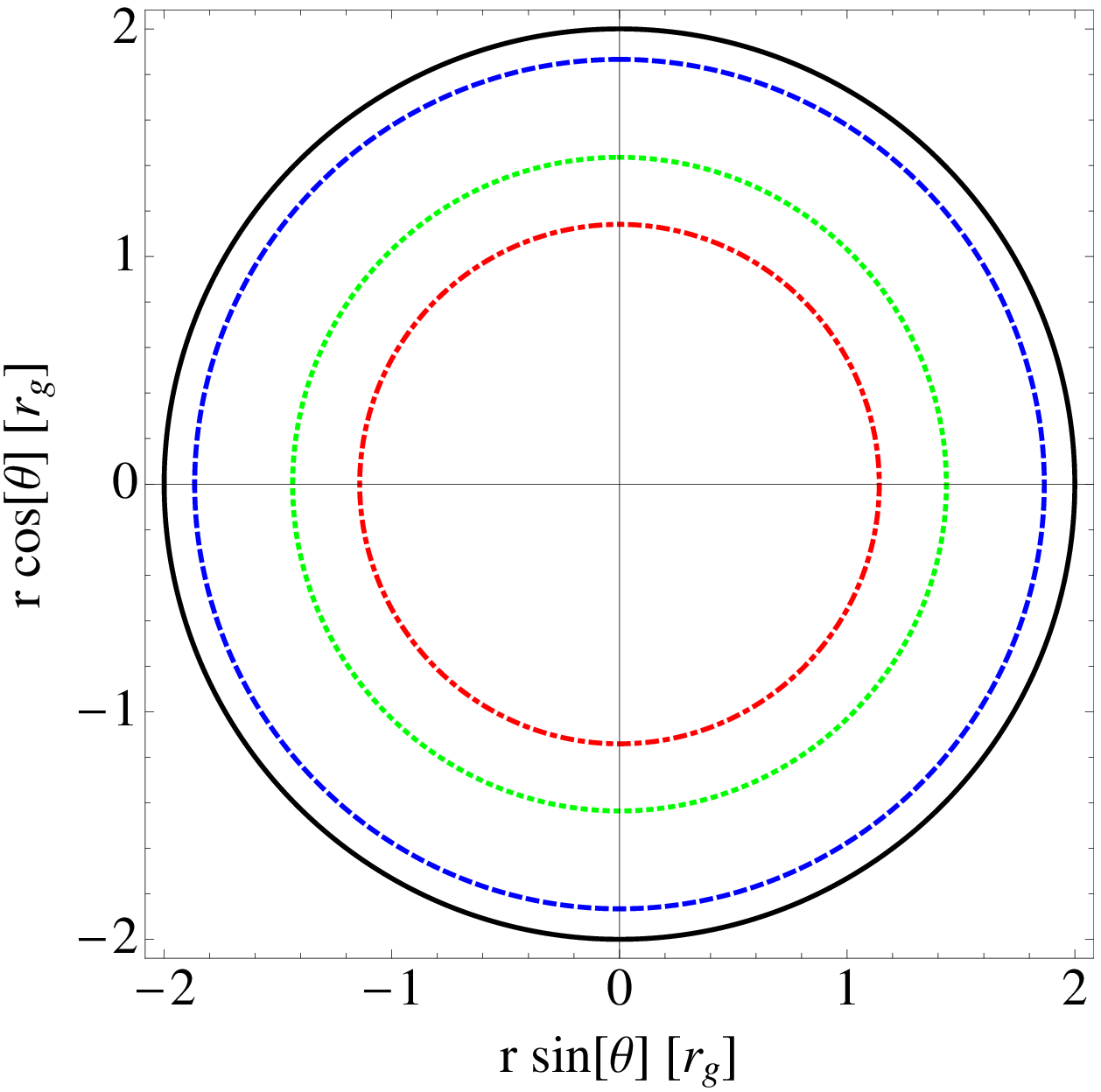}}
\end{minipage}
\begin{minipage}{8cm}
\resizebox{8cm}{!}{\plotone{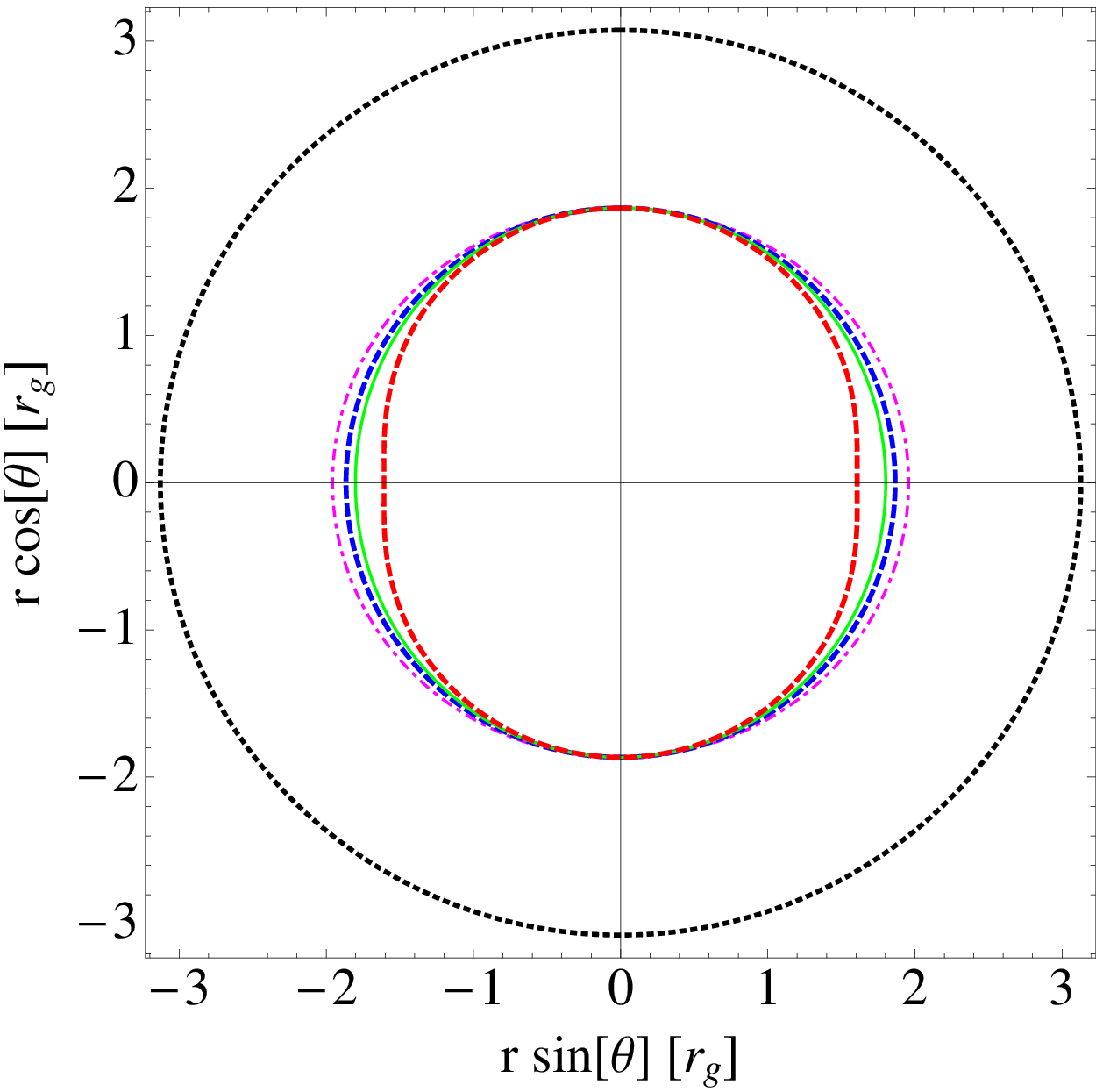}}
\end{minipage}
\begin{minipage}{8cm}
\resizebox{8cm}{!}{\plotone{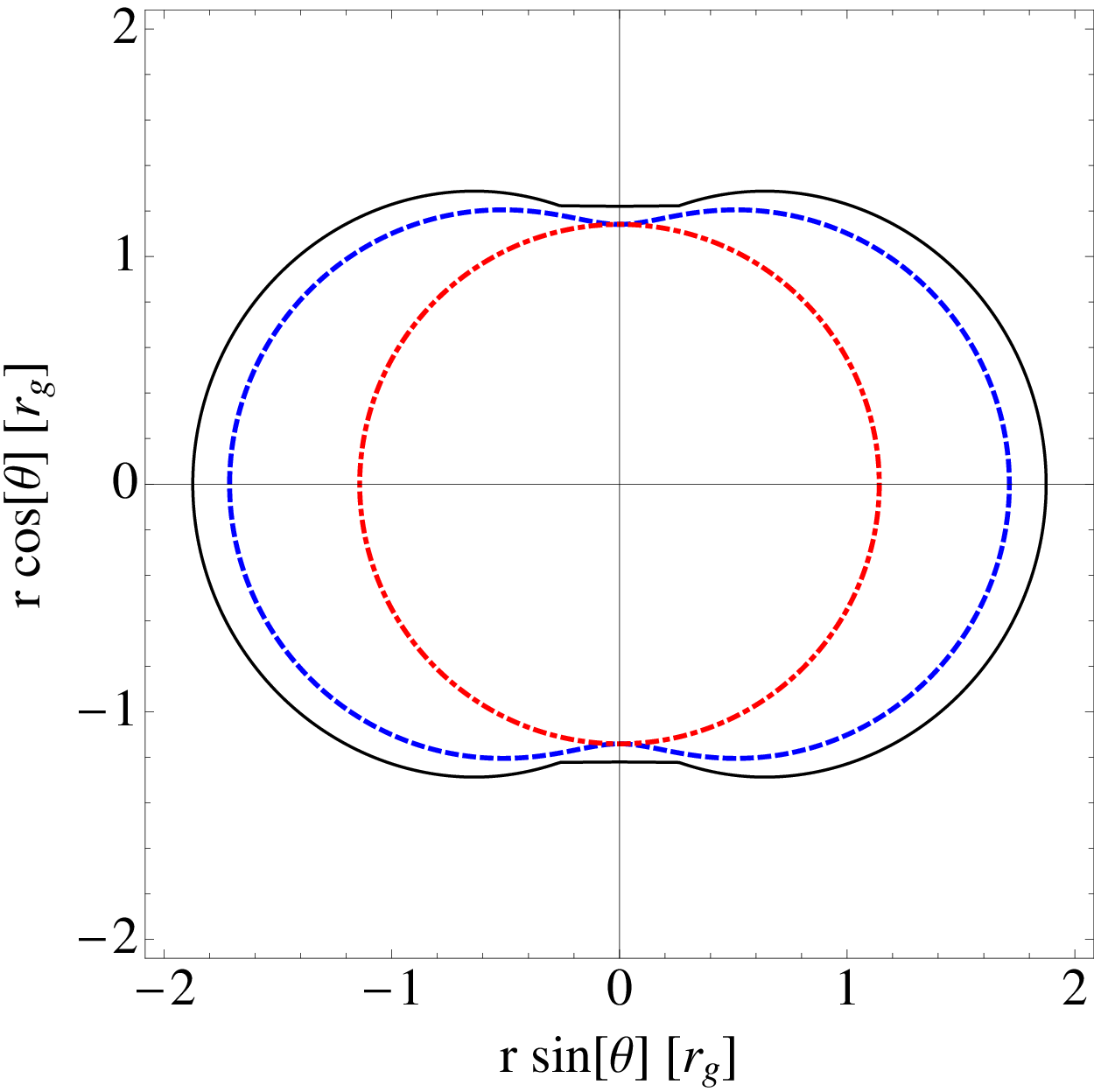}}
\end{minipage}

\caption{\label{F:eh} 
The outer event horizons of the considered black holes.
The upper left panel shows Kerr black holes with 
(from the outside inwards) $a=\,0,\,0.5\,,0.9,$ and 0.99;
the  upper right panel shows black holes with $a=0.5$ and
(from the outside inwards) $\epsilon_3\,=\,-30.6,\,-5,0\,,2.5$ and 6.3;
the lower panel shows black holes with $a=0.99$ and
(from the outside inwards) $\epsilon_3\,=\,-5,\,-2.5,$ and 0.}
\end{figure}

Figure \ref{F:eh} presents the event horizons in GC coordinates for all considered spacetimes. 
Note that the requirement of a closed horizon (and the absence of a naked singularity) limits the allowed range 
of $\epsilon_3$ to $\epsilon_3<\epsilon_{\rm max}$ with 
$\epsilon_{\rm max}=$6.75 for $a=0.5$, and $\epsilon_{\rm max}=$0.01 for  $a=0.99$.
For $a=0.99$ and  $\epsilon_3=$ -5, the event horizon shows a kink close to the pole 
where two roots of the equation defining the event horizons cross.
 
\begin{figure}
\resizebox{15cm}{!}{\plotone{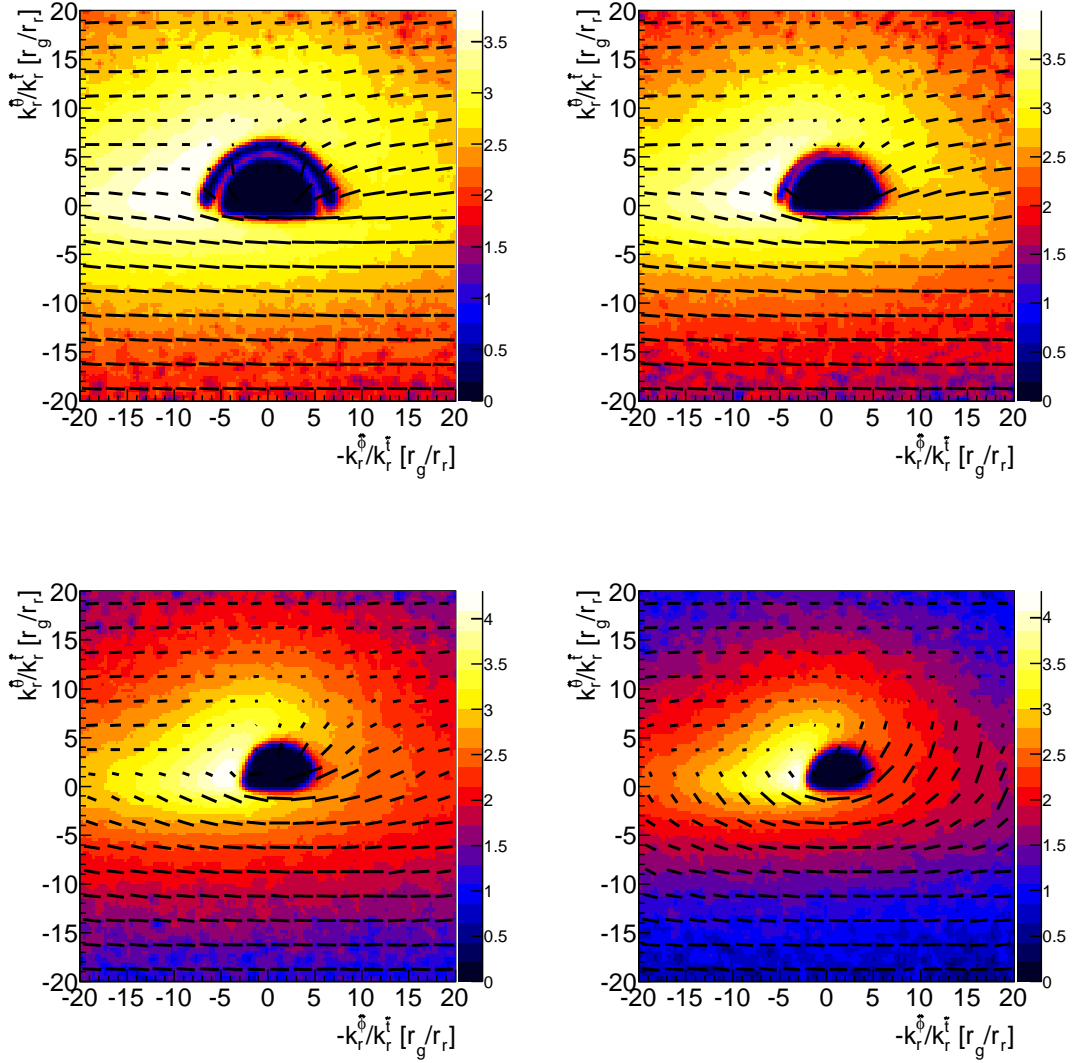}}
\caption{\label{F:img1} Simulated images (0.1-10 keV) of four accreting Kerr black holes for an inclination of 75$^{\circ}$. 
The results are shown for 
$a=0$ (top-left, Model A), 
$a=0.5$ (top-right, Model B), 
$a=0.9$ (bottom-left, Model C), and
$a=0.99$ (bottom-right, Model D).
See Table \ref{T:metric} for all the parameters of the different models.
The logarithmic color scale shows the 0.1-10 keV photon number flux. 
The length of the bars show the polarization degree 
and the orientation shows the polarization direction of the electric field vector. 
Far away from the black hole the polarization degree is $\sim$4\%.}
\end{figure}

\begin{figure}
\resizebox{15cm}{!}{\plotone{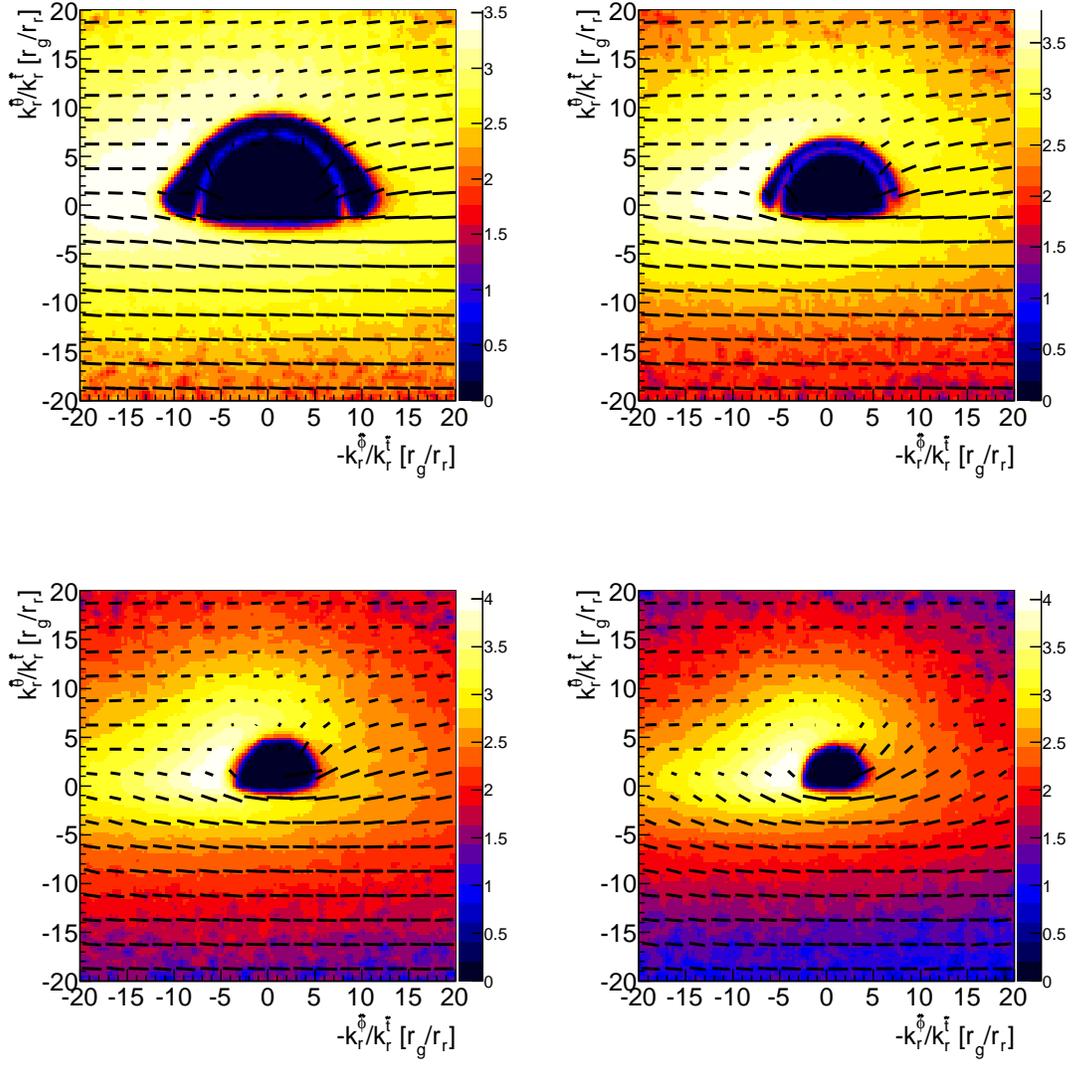}}
\caption{\label{F:img2} Same as Fig.\ \ref{F:img1}, but for  
$a=0.5$ and $\epsilon_3=-30.6$ (top-left, Model E),
$a=0.5$ and $\epsilon_3=-5$ (top-right, Model F),
$a=0.5$ and $\epsilon_3=2.5$ (bottom-left, Model G), and
$a=0.5$ and $\epsilon_3=6.3$ (bottom-right, Model H).}
\end{figure}

\begin{figure}
\resizebox{15cm}{!}{\plotone{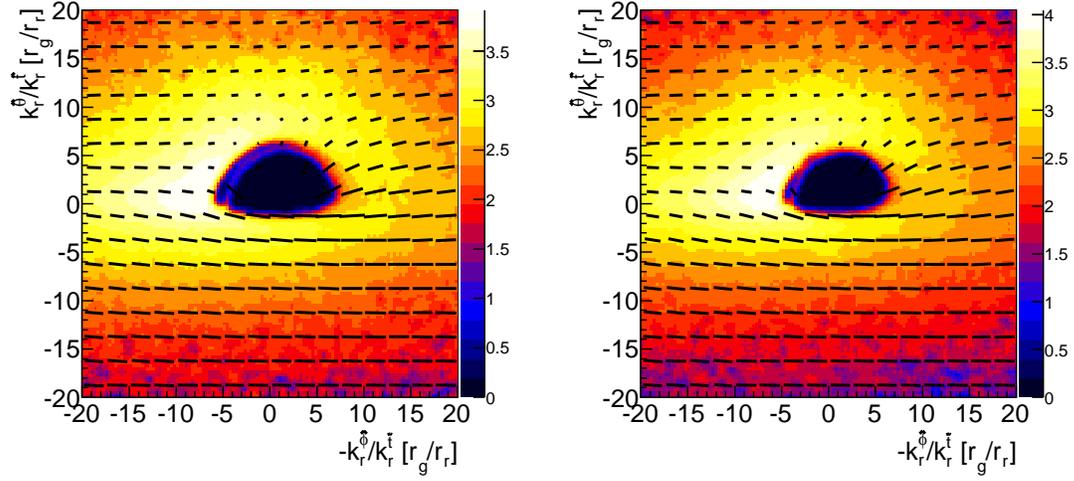}}
\caption{\label{F:img3} Same as Fig.\ \ref{F:img1}, but for $a=0.99$ and $\epsilon_3=-5$ (left panel, Model I)
and $a=0.99$ and $\epsilon_3=-2.5$ (right panel, Model J).}
\end{figure}

Figures \ref{F:img1}-\ref{F:img3} show the ray tracing images of the accretion disks for all simulated models
for an inclination of 75$^{\circ}$ from the rotation axis.
The lengths and orientations of the superimposed lines show the polarization degree and polarization direction, 
respectively.  The brightness maps exhibit the well-known asymmetric appearance resulting from the relativistic beaming 
and de-beaming of the emission from the disk material approaching and receding from the observer with a 
velocity close to the speed of light, respectively, and the asymmetric gravitational lensing in the curved spacetime.
The rings between the event horizon and the inner edge of the accretion disk come from photons orbiting 
around the black hole for multiples of $\sim 180^{\circ}$. The brightness maps of the GR and non-GR models look 
somewhat similar. The largest differences result from a larger (smaller) ISCO of the non-GR models 
for negative (positive) $\epsilon_3$-values. 
It is instructive to compare model A ($\epsilon_3=0$, $a=0$, Fig.\ \ref{F:img1}, upper-left panel) with model I
($\epsilon_3=-5$, $a=0.99$,  Fig.\ \ref{F:img3}, left panel): although the images of the ISCOs have similar 
angular diameters, model I shows the distortions that are typical for high spin values.  
From a theoretical viewpoint, the parameter $M$ is the mass of the black hole as measured 
by a distant observer and is thus not a free parameter; the two images thus demonstrate that 
the metric from Equ.\ (\ref{E:metric}) with $\epsilon_3\ne 0$ indeed leads to new observable 
characteristics, and does not merely correspond to the Kerr metric in different coordinates.
Note that in practice $M$ may be known from the observations of the companion star,  see e.g.\ 
\citep{Oro:11}.
For models with rather small ISCOs (Models D and H), the polarization direction of the emission from 
the inner disk exhibits a pattern that ``circles around the black hole'' owing to the dominance of the scattered emission.
For the other models, the combination of GR effects and the competition between the direct and
scattered emission results in a complex polarization pattern with subtle differences 
between the GR and the non-GR models.  

Figures \ref{F:C0}-\ref{F:C2} present the accretion disk brightness $F(r)$, the energy spectrum,
the polarization degree spectrum, and the polarization direction spectrum for all simulated models, again
for an inclination of 75$^{\circ}$ from the rotation axis. 
In the case of the Kerr black holes with fixed mass $M$ (Fig.\ \ref{F:C0}), 
$a$ determines the location of the event horizon and the ISCO.
With increasing $a$, the ISCO moves towards smaller $r$,
the disk brightness and disk temperature increase close to the black hole
and the energy spectra extend to higher energies.
The energy spectrum of the scattered radiation is harder than that of the direct emission
owing to the facts that the scattered emission originates from the hot inner accretion disk and that
the scattering can increase the photon energy owing to an additional Doppler boost.
As a consequence of the hard spectrum and the relatively high polarization degree 
(compared to the direct emission), the scattered emission strongly impacts the 
polarization properties at higher energies.
The polarization direction exhibits a swing from horizontal polarization at low energies 
(owing to the emission of the optically thick accretion disk) to vertical polarization 
at high energies (owing to scattered emission). 
The competition of the horizontal and vertical polarizations leads to a minimum of the polarization degree
at intermediate energies. The swing of the polarization direction and the minimum of the polarization 
degree are observed at lower energies for the smaller ISCO models owing to the increased overall 
importance of scattered radiation. 

The corresponding results are shown in Figs.\ \ref{F:C1} and \ref{F:C2} for the non-GR black holes 
with $a$=0.5 and $a$=0.99, respectively. The results exemplify that the observational results 
strongly depend on the location of the ISCO: the smaller $\epsilon_3$, the smaller $r_{\rm ISCO}$, the harder 
the energy spectra, and the more pronounced is the energy dependence of the polarization properties.

Under the assumptions made in this paper, GR and non-GR models produce qualitatively different
energy spectra and polarization spectra for $\epsilon_3$-values which lead to $r_{\rm ISCO}>6\,r_{\rm g}$. 
As an example, Model E (with $a=0.5$, $\epsilon_3\,=\,-30.6$, $r_{\rm ISCO}\,=\,10\,r_{\rm g}$) exhibits a softer 
energy spectrum and less variation of the polarization degree and polarization direction than any of the Kerr-models. 
Existing X-ray spectroscopic data can already be used to exclude such negative $\epsilon_3$-values. 

The difference between GR and non-GR models is rather small for models with the same $r_{\rm ISCO}$.
As an example, Fig.\ \ref{F:C3} shows GR and non-GR models with approximately the same $r_{\rm ISCO}$-values. 
The GR and non-GR models produce almost indistinguishable energy spectra and similar - but not identical - 
polarization properties. The differences between the GR and non-GR models are somewhat larger for 
rapidly spinning black holes because more photons are emitted and propagate close to the event 
horizon where the JP11 metric deviates most strongly from the Kerr metric.
Although the GR model C and the non-GR model H have the same ISCO and show almost identical flux energy spectra, 
they exhibit different polarization energy spectra. The latter depend more strongly on the underlying 
spacetime as the competition of the direct and scattered emission leads to more pronounced 
observational signatures owing to the very different polarization properties of the two 
emission components.
\section{Summary and Discussion}
\label{S:Discussion}
In this paper we explore the possibility to test GR in the strong gravity regime with X-ray spectropolarimetric 
observations of  black holes in X-ray binaries. In the thermal state, the accretions disks, and the
emission and scattering geometry and processes are relatively simple and well understood, 
making these systems attractive for the test of the underlying spacetime.
We developed a code to simulate the polarized emission of the accretion disk. The code
computes the radial structure and radial brightness profile of the accretion disk for
GR and non-GR spacetimes. Furthermore, it parallel transports the wave vector and the 
polarization vector and accounts for scattering based on the classical results of 
Chandrasekhar (C60).  

We used the code to study the observational differences between Kerr black holes and
the black holes described by the phenomenological metric of JP11. 
The main effect of the JP11 metric for X-ray spectral and polarization measurements is to allow 
for other ISCOs than those predicted by GR for a certain $M$ and $a$ combination. 
The X-ray spectropolarimetric observations will allow us to measure allowed and 
forbidden regions in the plane of the parameters $a$ (black hole spin) and $\epsilon_3$ 
(quantifying the deviation from the Kerr metric).
However, for large regions of the parameter space, the approximate degeneracy between the 
black hole spin $a$ and the deviation parameter $\epsilon_3$ will make it difficult to 
distinguish between GR and non-GR models.
  
\begin{figure}
\epsscale{.80}
\begin{minipage}{8cm}
\resizebox{8cm}{!}{\plotone{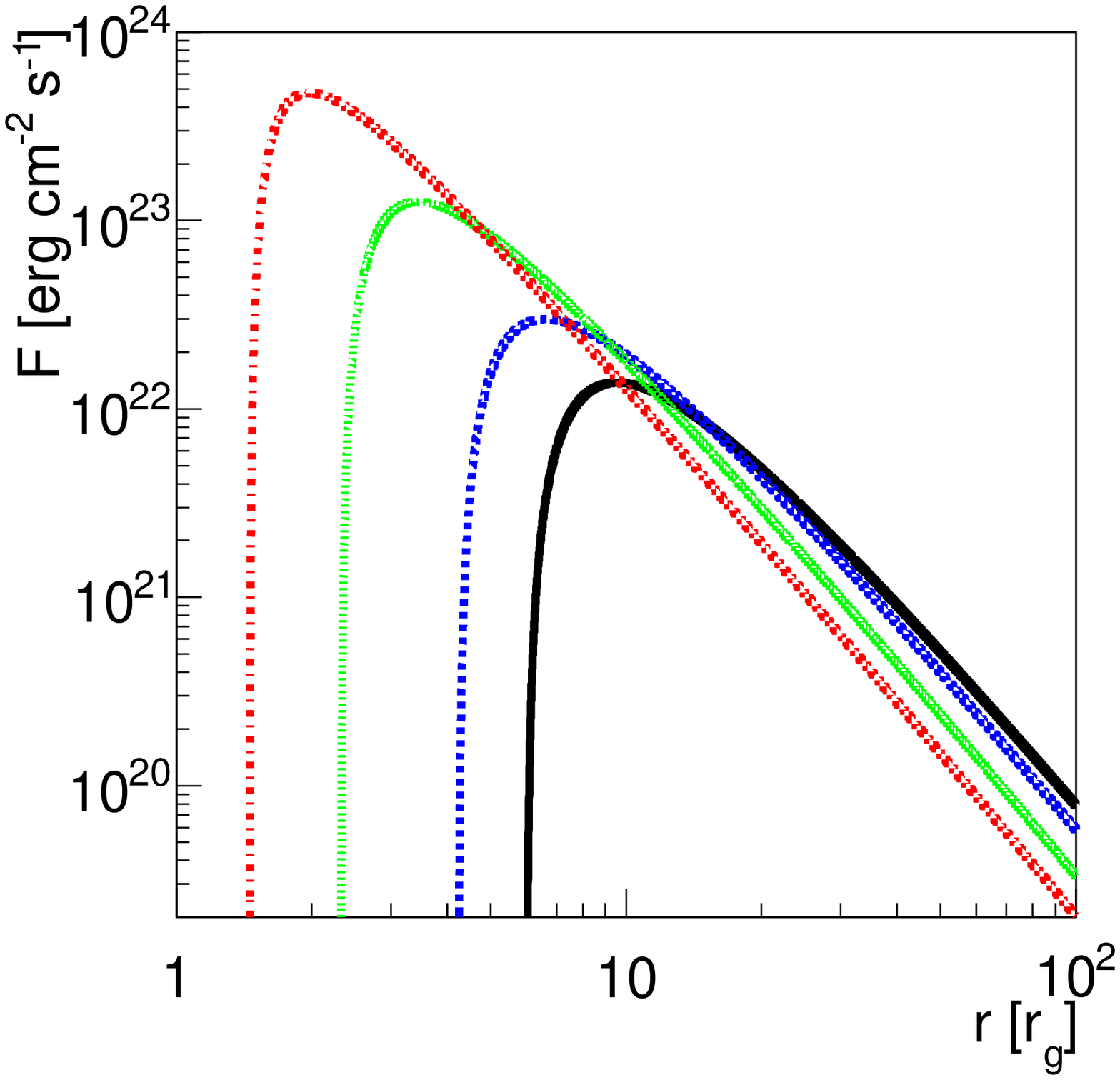}}
\end{minipage}
\begin{minipage}{8cm}
\resizebox{8cm}{!}{\plotone{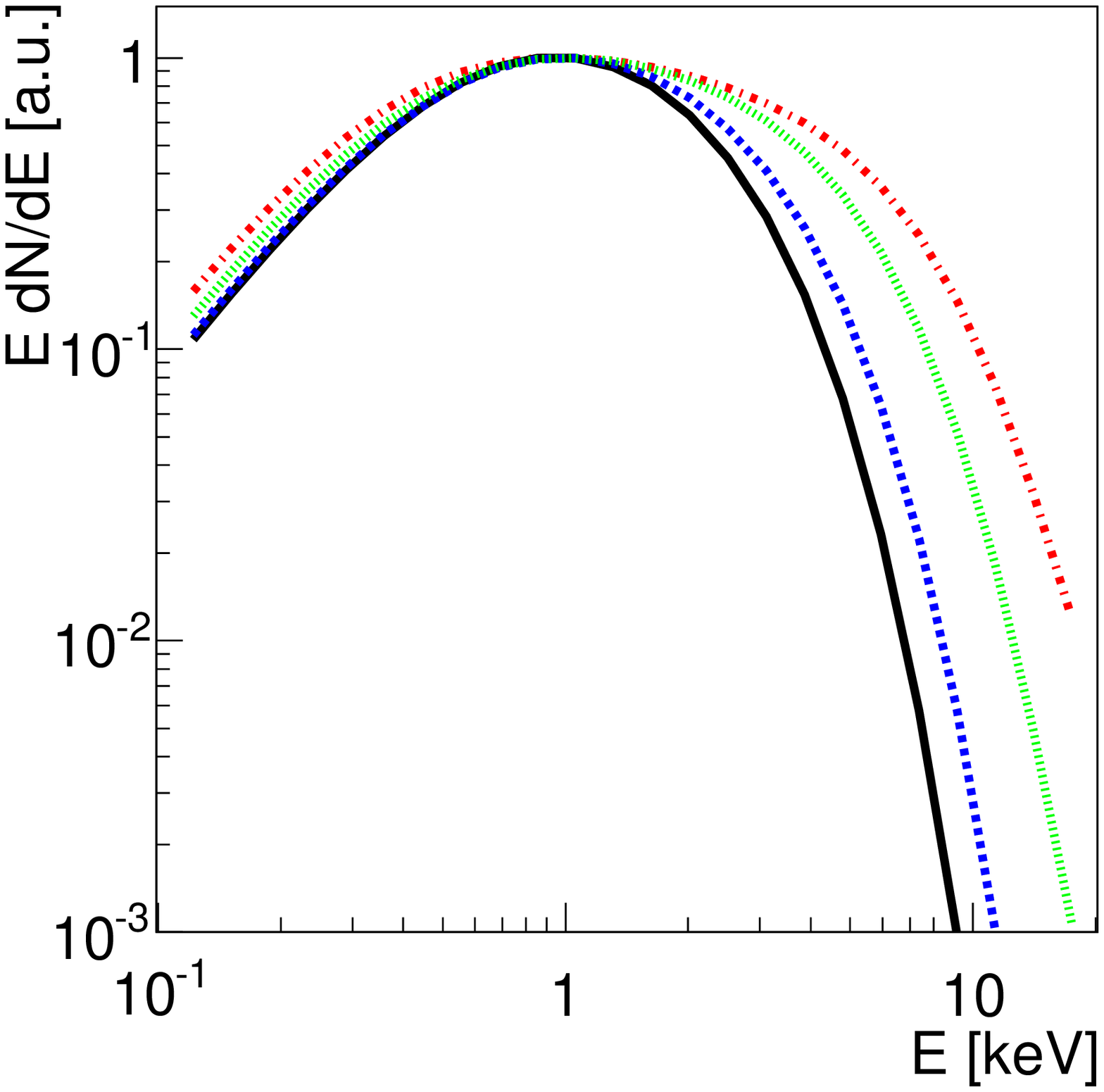}}
\end{minipage}
\begin{minipage}{8cm}
\resizebox{8cm}{!}{\plotone{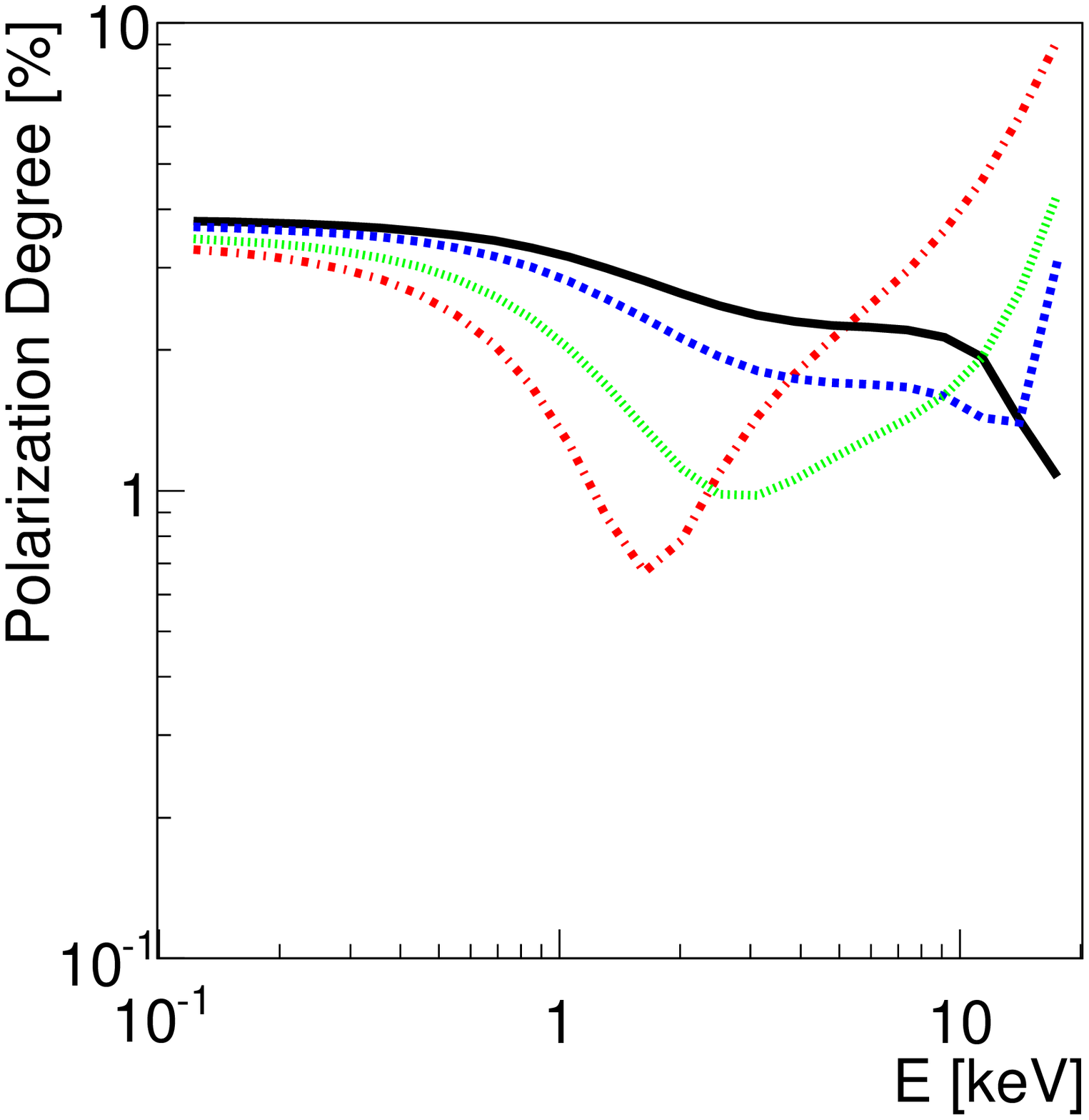}}
\end{minipage}
\begin{minipage}{8cm}
\resizebox{8cm}{!}{\plotone{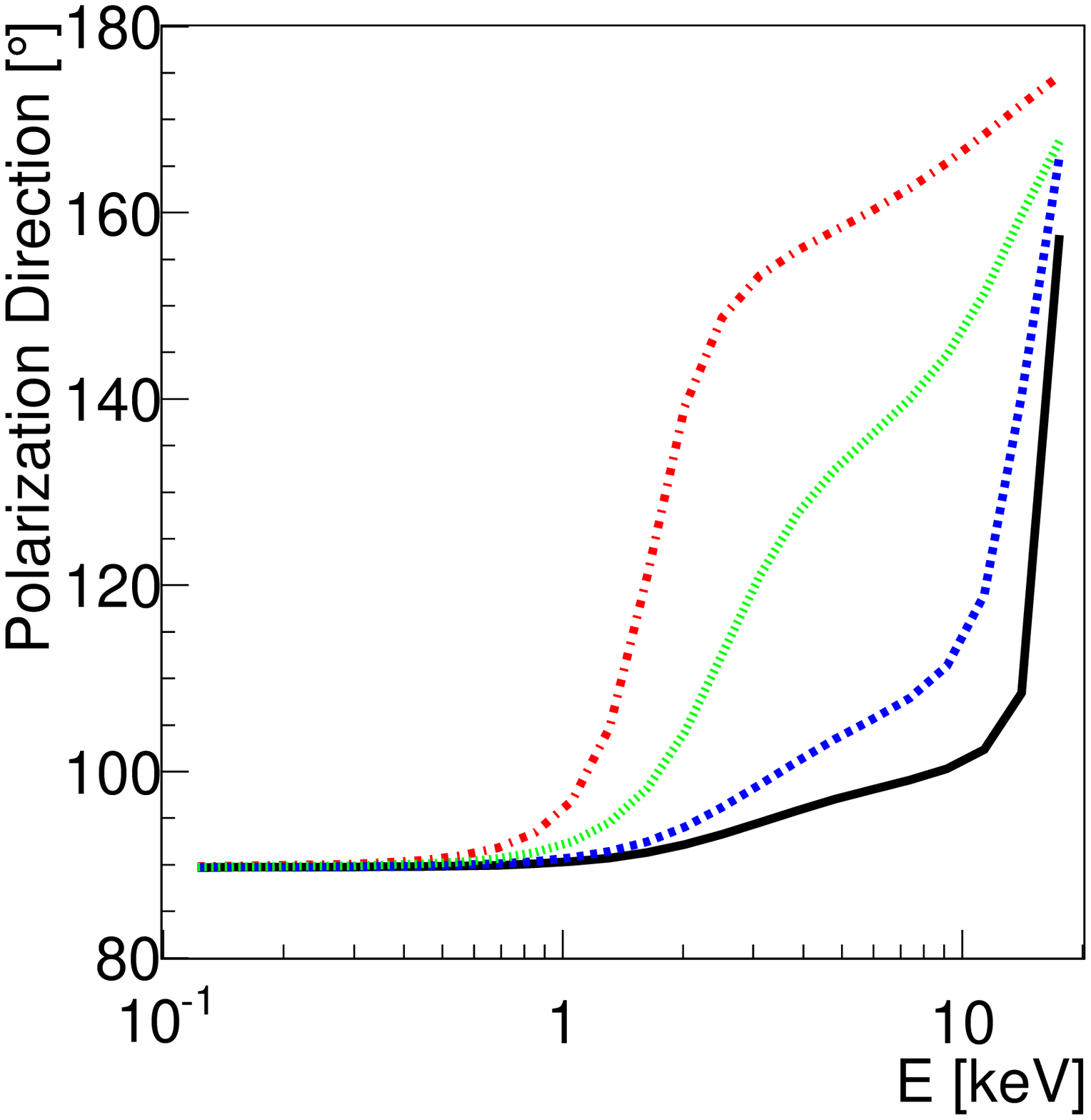}}
\end{minipage}
\caption{\label{F:C0} Accretion disk brightness in the plasma frame (top-left), energy spectrum (top-right),
polarization degree (bottom-left) and polarization direction (bottom-right) for four Kerr black holes with
$a=0$ (solid black line, Model A),
$a=0.5$ (dashed blue line, Model B),
$a=0.9$ (dotted green line, Model C), and
$a=0.99$ (dash-dotted red line, Model D).
The inclination is 75$^{\circ}$. The polarization direction $\chi$ is measured from the projection 
of the spin axis of the black hole in the sky with
$\chi$ increasing for a clockwise rotation of the polarization direction when looking into the beam.  
}
\end{figure}

\begin{figure}
\epsscale{.80}
\begin{minipage}{8cm}
\resizebox{8cm}{!}{\plotone{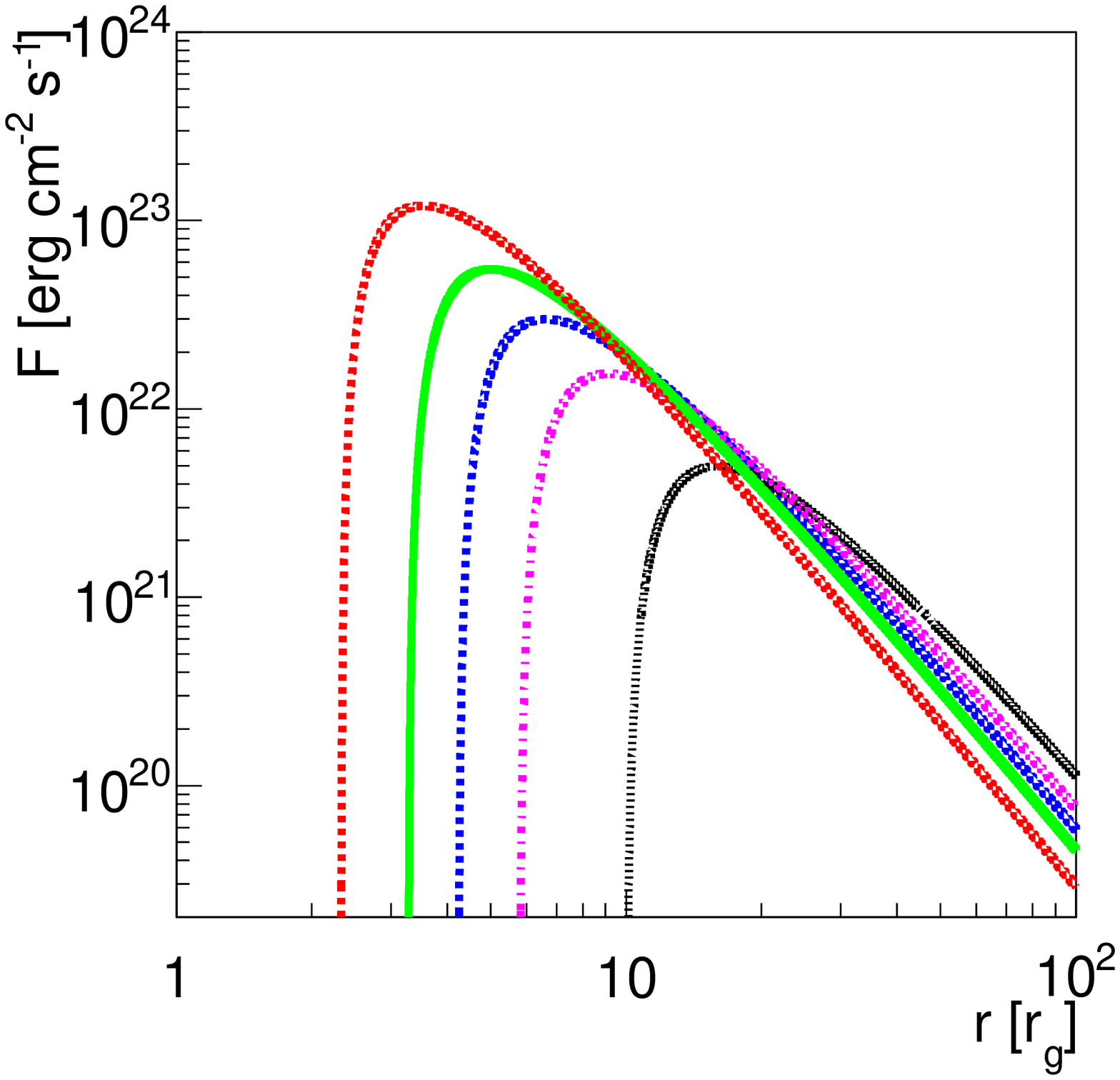}}
\end{minipage}
\begin{minipage}{8cm}
\resizebox{8cm}{!}{\plotone{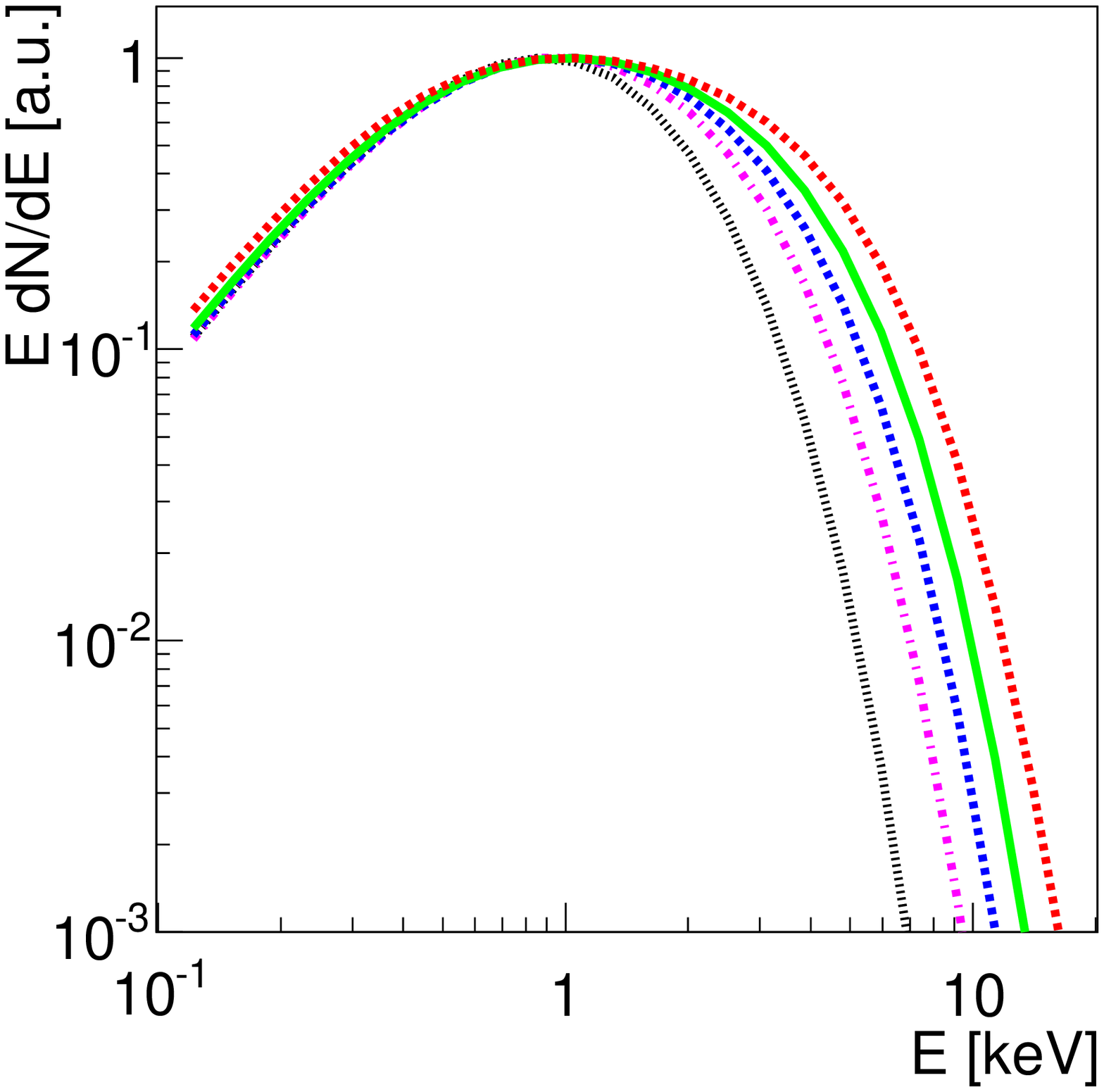}}
\end{minipage}
\begin{minipage}{8cm}
\resizebox{8cm}{!}{\plotone{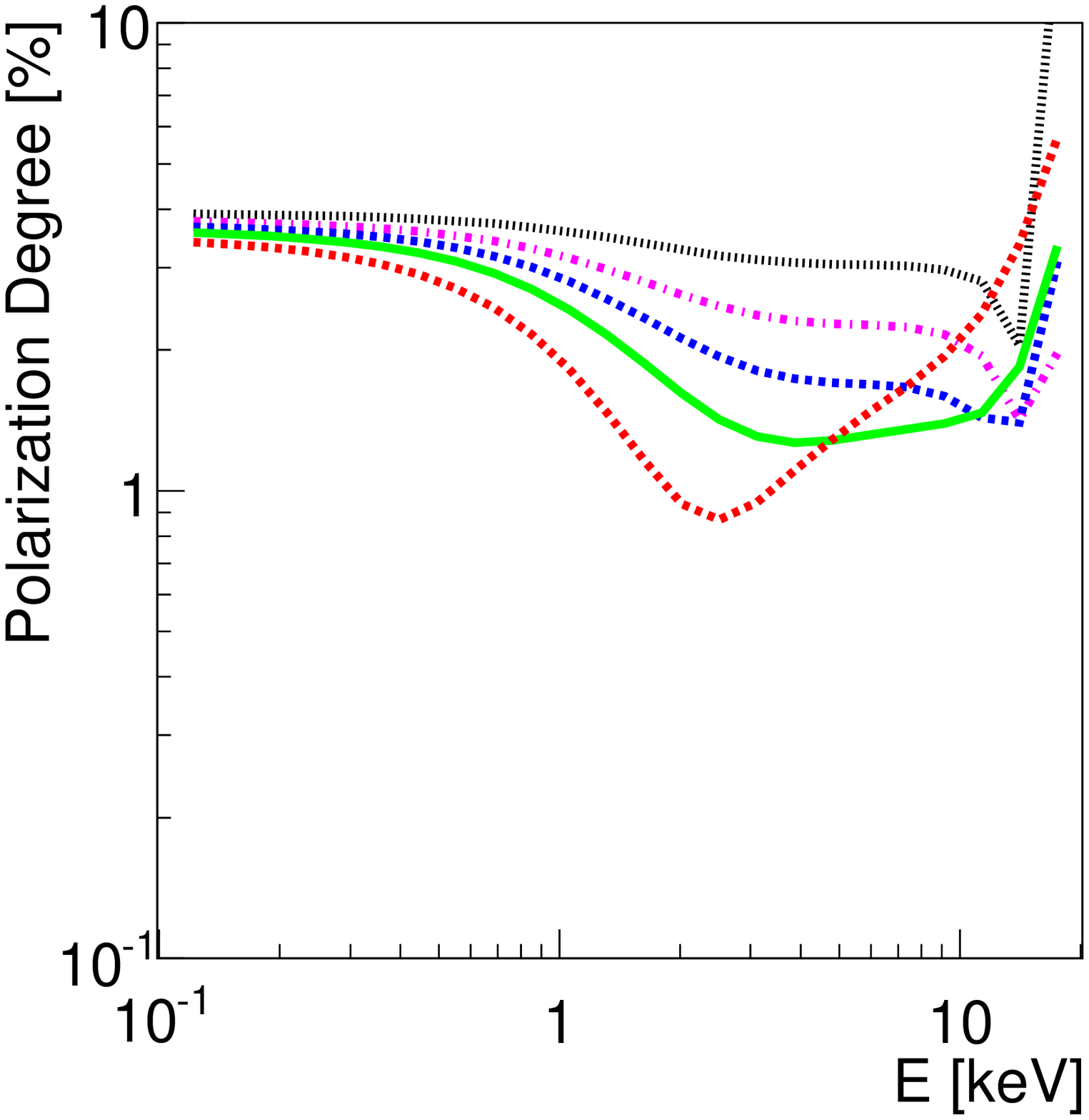}}
\end{minipage}
\begin{minipage}{8cm}
\resizebox{8cm}{!}{\plotone{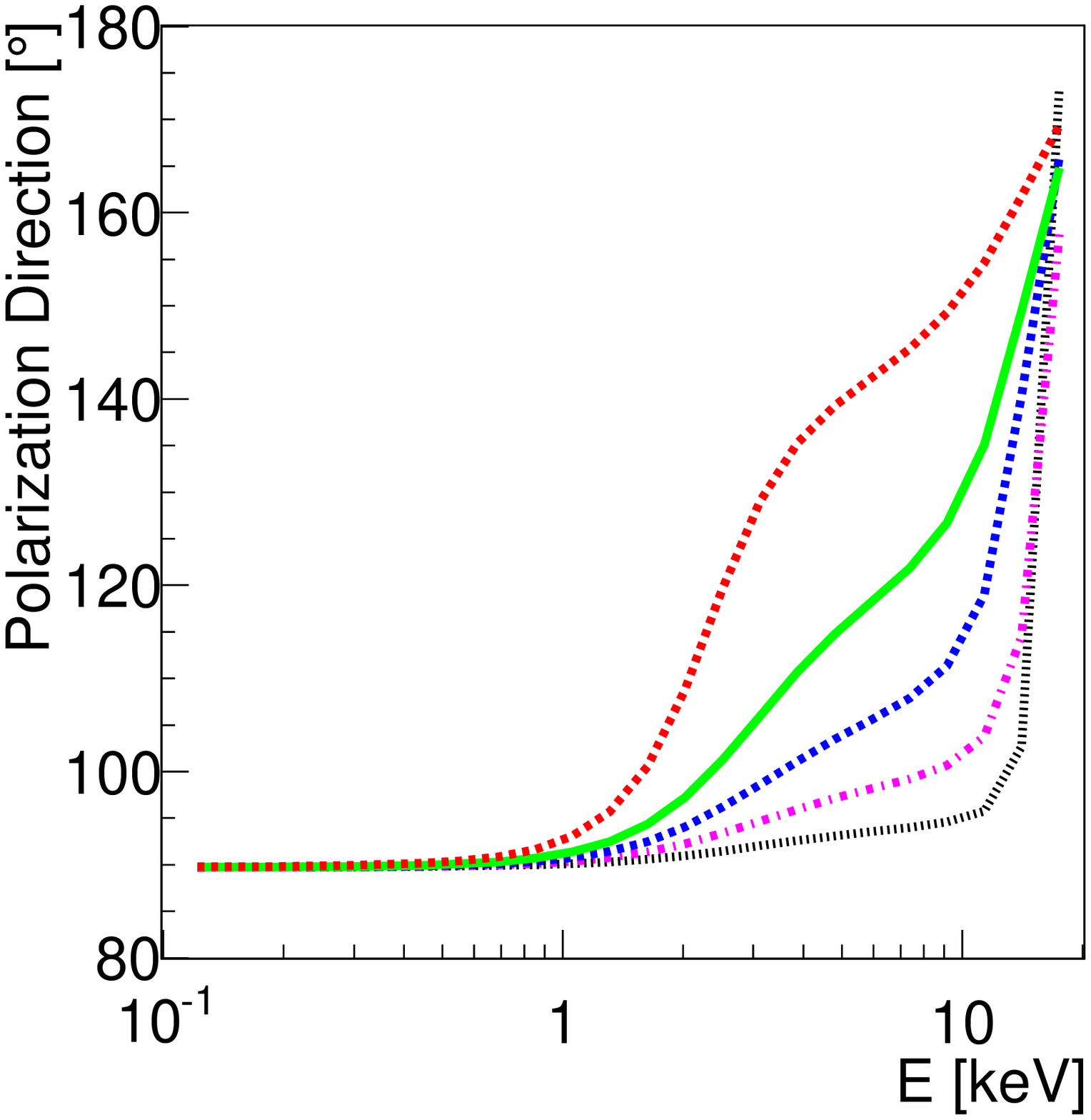}}
\end{minipage}
\caption{\label{F:C1} The same as Fig.\ \ref{F:C0} but for black holes with 
$a=0.5$ and $\epsilon_3=-30.6$ (dotted black line, Model E),
$a=0.5$ and $\epsilon_3=-5$ (dash-dotted magenta line, Model F),
$a=0.5$ and $\epsilon_3=0$ (dashed blue line, Model B),
$a=0.5$ and $\epsilon_3=2.5$ (solid green line, Model G), and
$a=0.5$ and $\epsilon_3=6.3$ (dashed red line, Model H).}
\end{figure}

\begin{figure}
\epsscale{.80}
\begin{minipage}{8cm}
\resizebox{8cm}{!}{\plotone{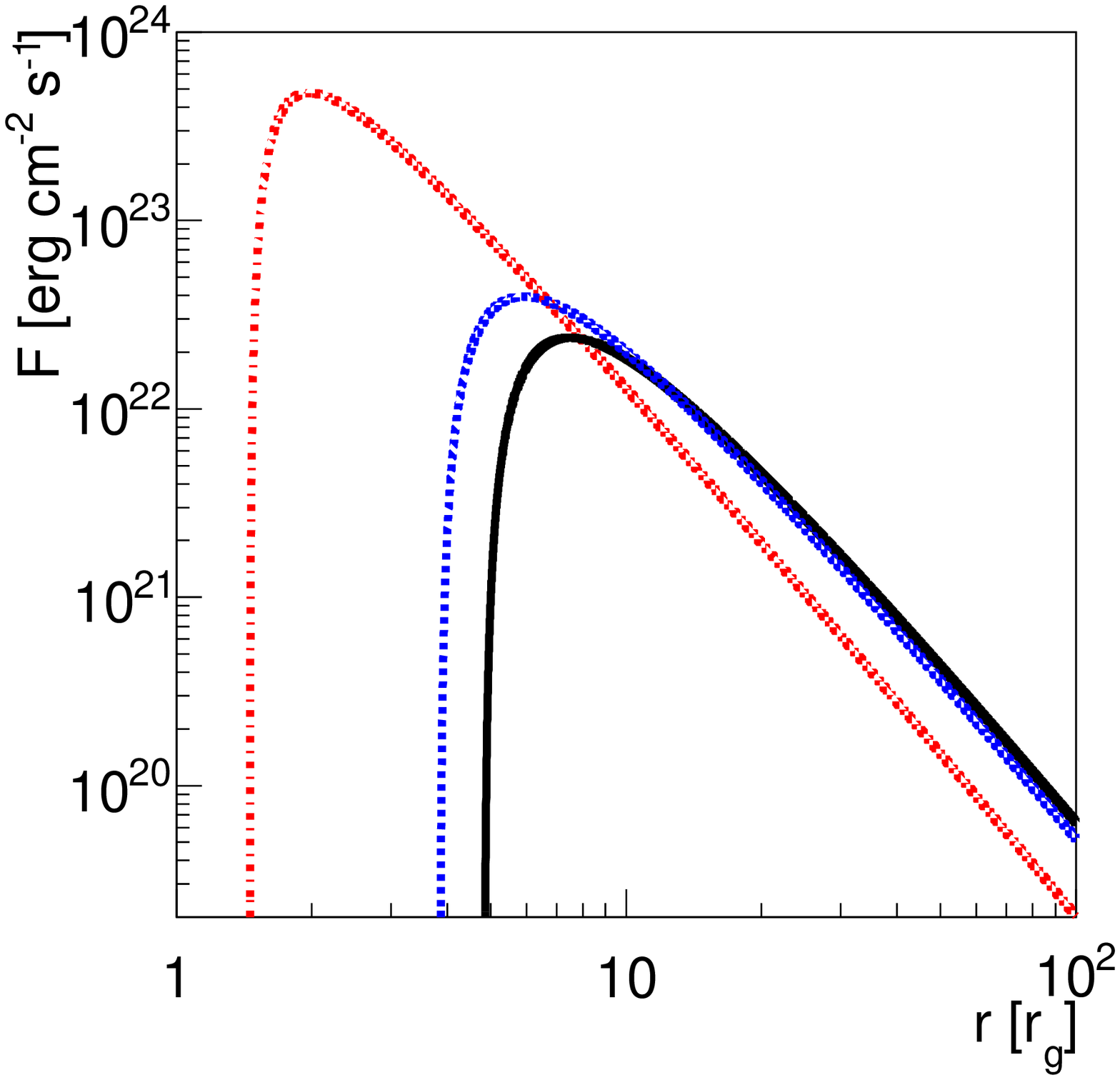}}
\end{minipage}
\begin{minipage}{8cm}
\resizebox{8cm}{!}{\plotone{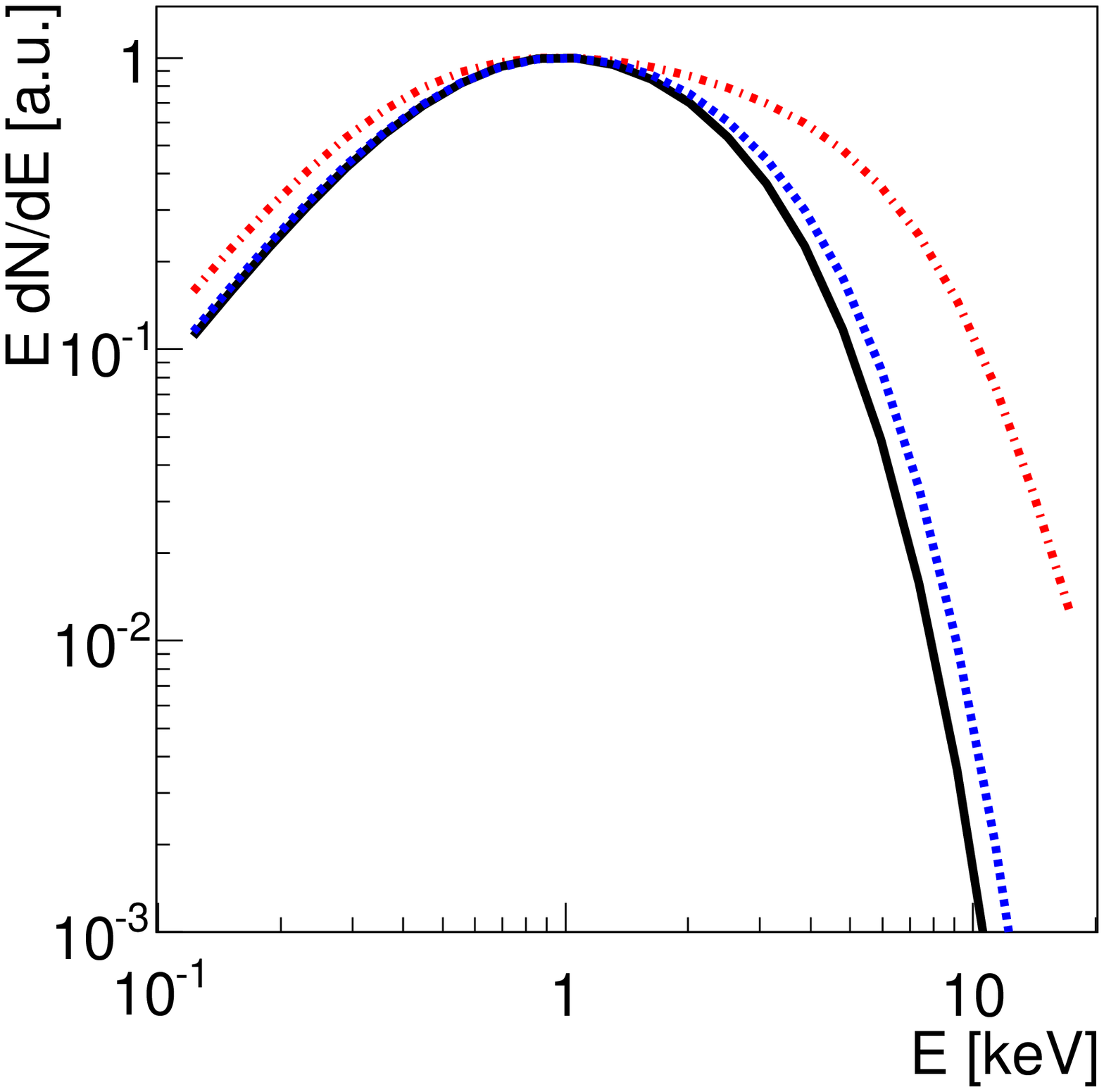}}
\end{minipage}
\begin{minipage}{8cm}
\resizebox{8cm}{!}{\plotone{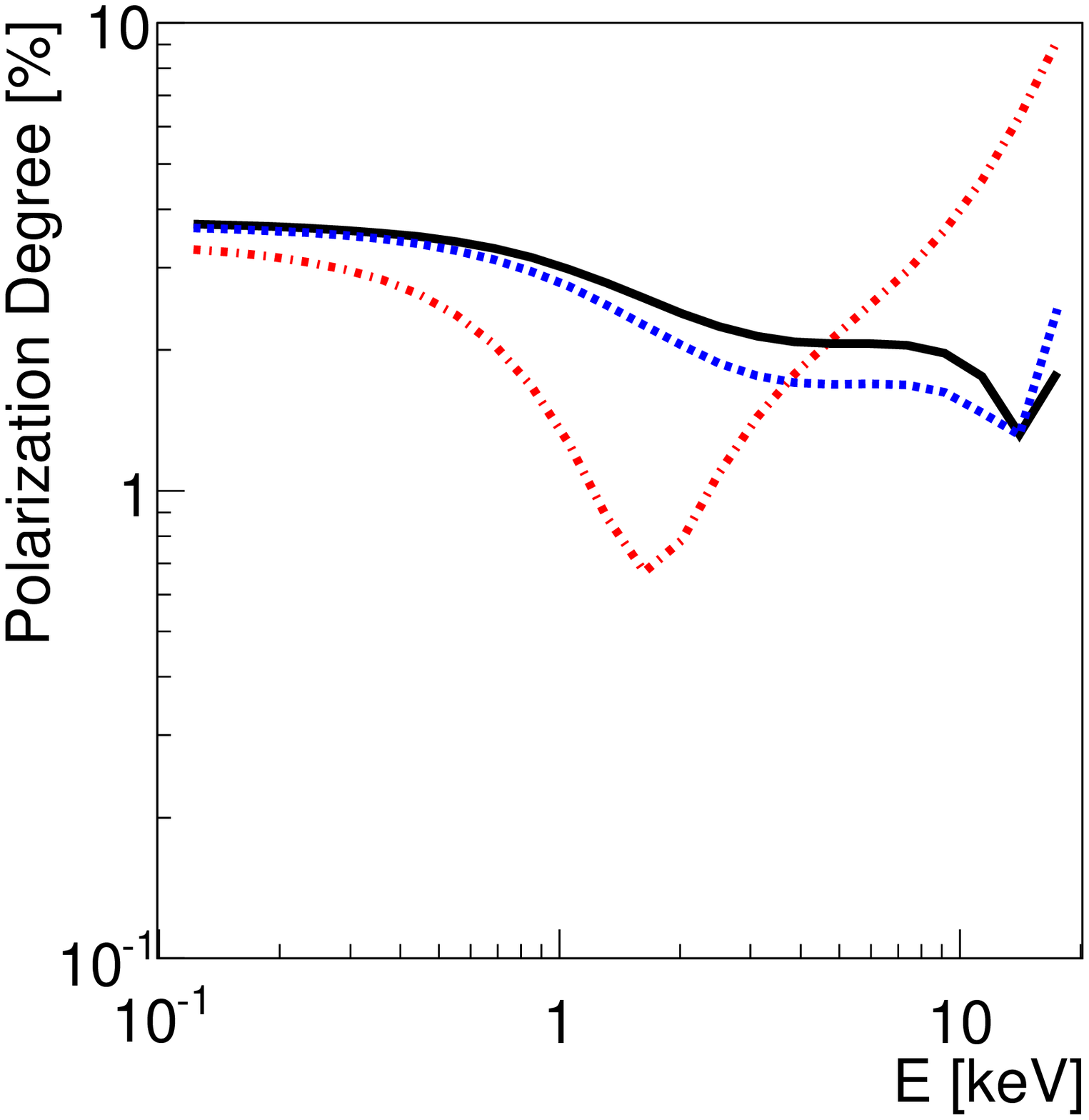}}
\end{minipage}
\begin{minipage}{8cm}
\resizebox{8cm}{!}{\plotone{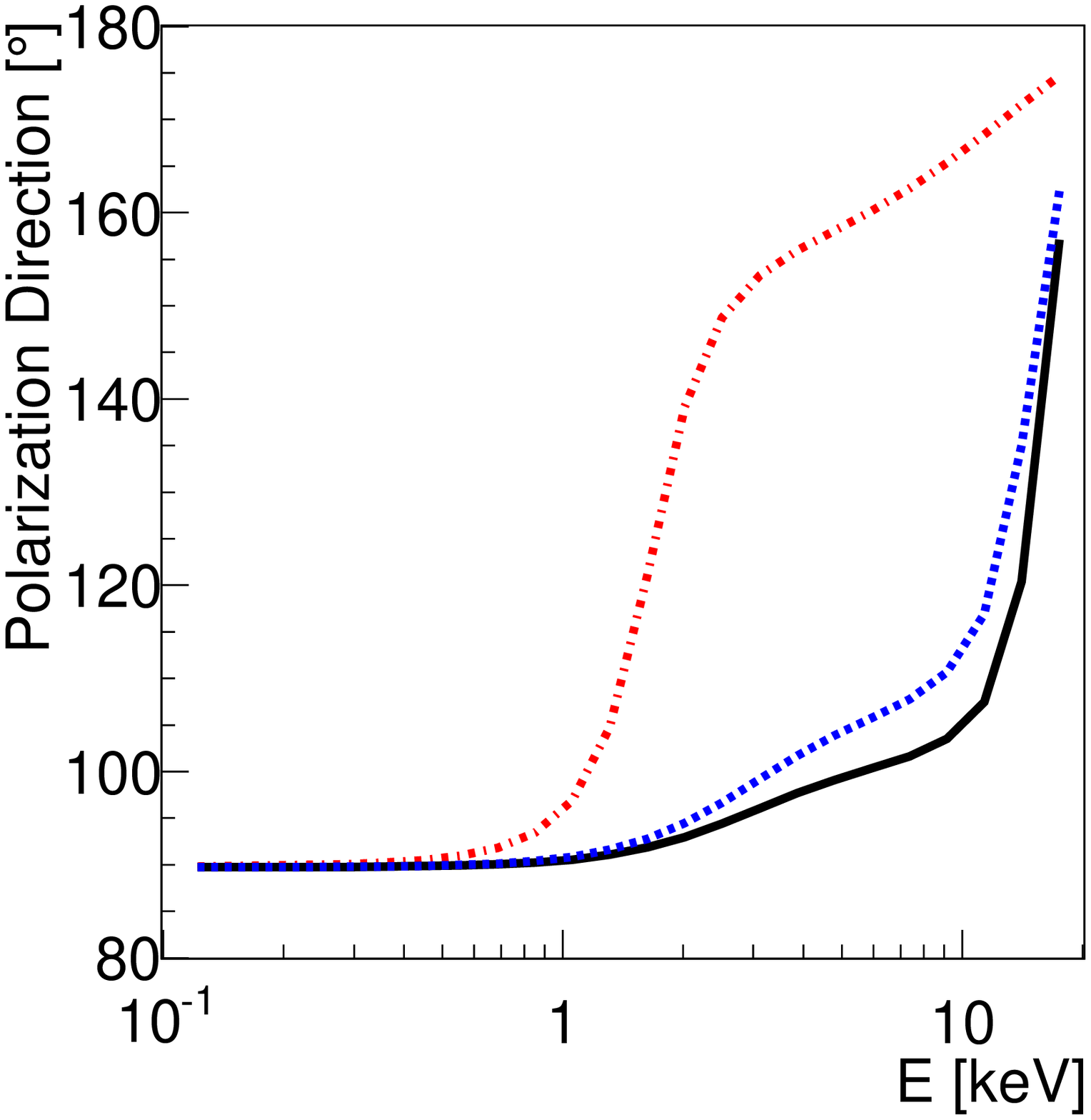}}
\end{minipage}
\caption{\label{F:C2} The same as Fig.\ \ref{F:C0} but for black holes with 
$a=0.99$ and $\epsilon_3=-5$ (solid black line, Model I),
$a=0.99$ and $\epsilon_3=-2.5$ (dashed blue line, Model J), and
$a=0.99$ and $\epsilon_3=0$ (dash-dotted red line, Model D). }
\end{figure}

\begin{figure}
\epsscale{.80}
\begin{minipage}{8cm}
\resizebox{8cm}{!}{\plotone{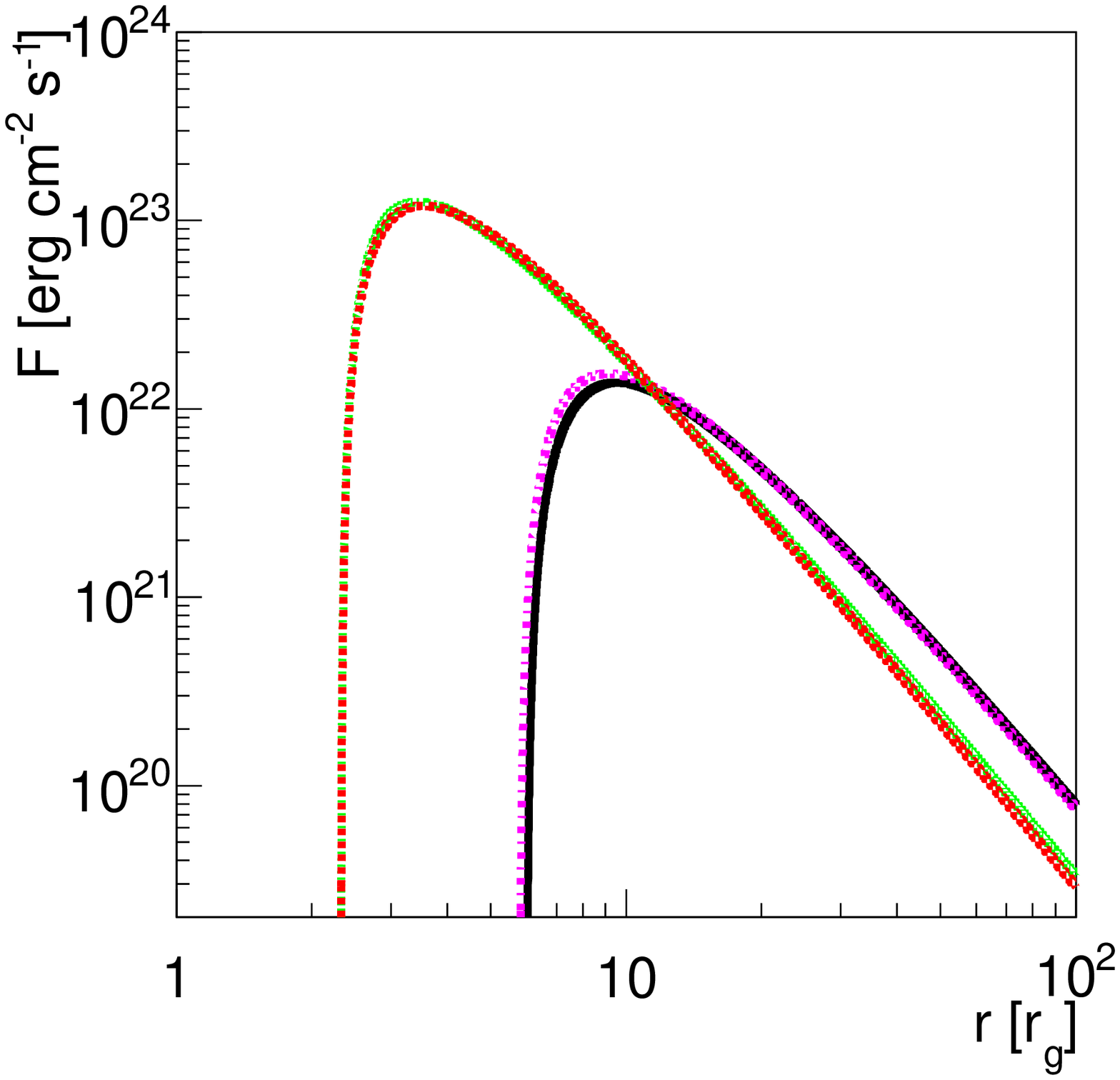}}
\end{minipage}
\begin{minipage}{8cm}
\resizebox{8cm}{!}{\plotone{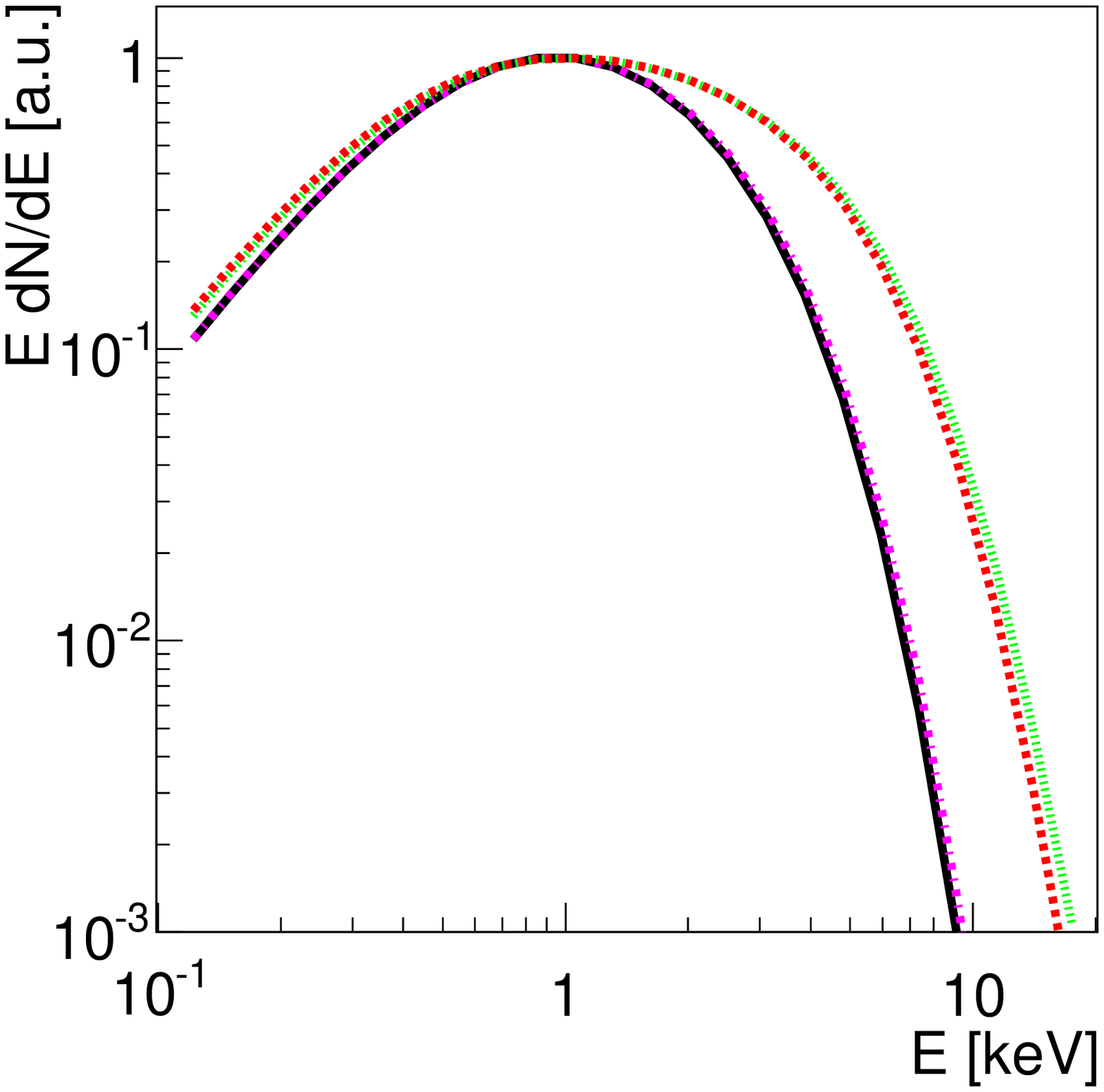}}
\end{minipage}
\begin{minipage}{8cm}
\resizebox{8cm}{!}{\plotone{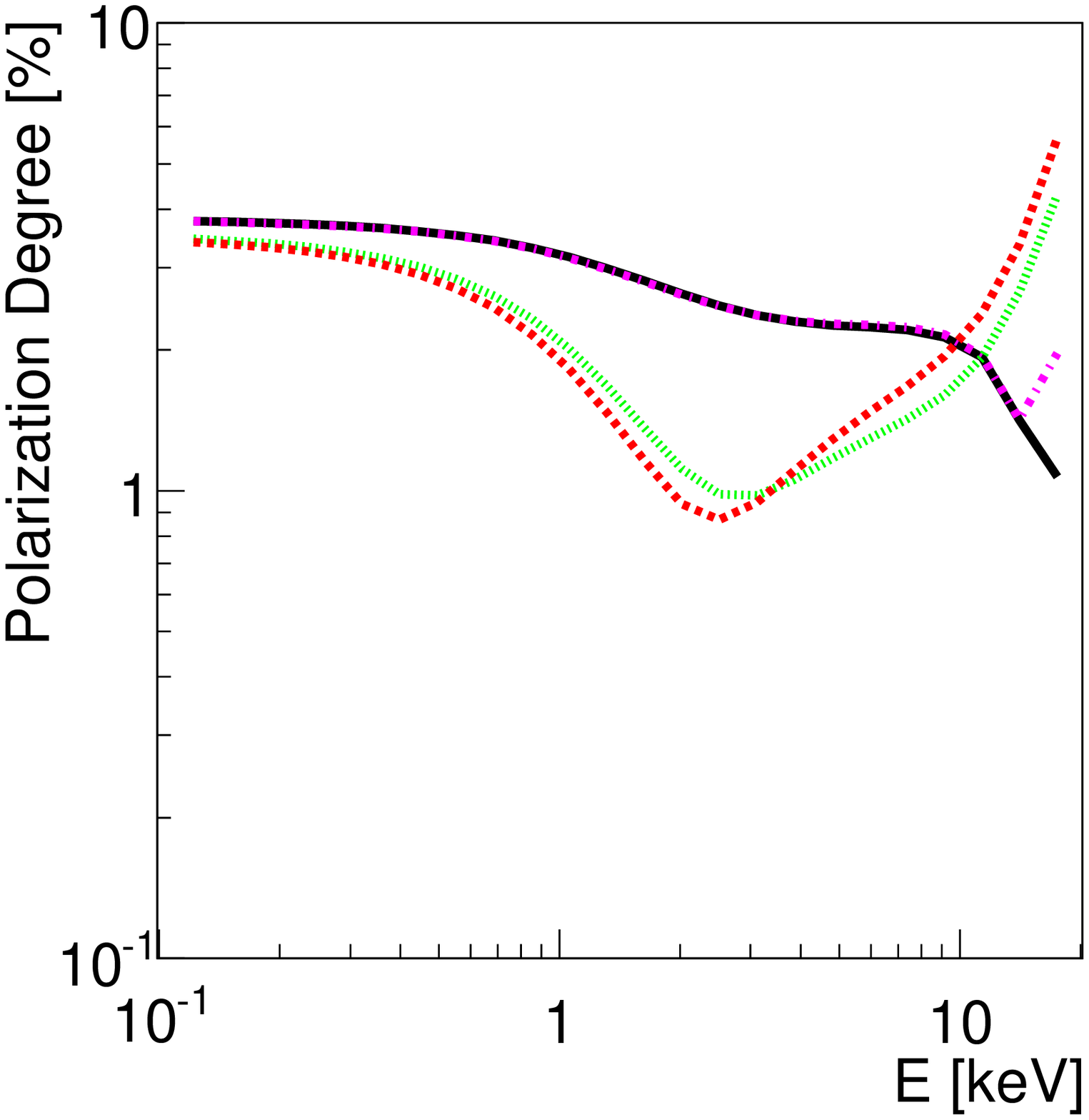}}
\end{minipage}
\begin{minipage}{8cm}
\resizebox{8cm}{!}{\plotone{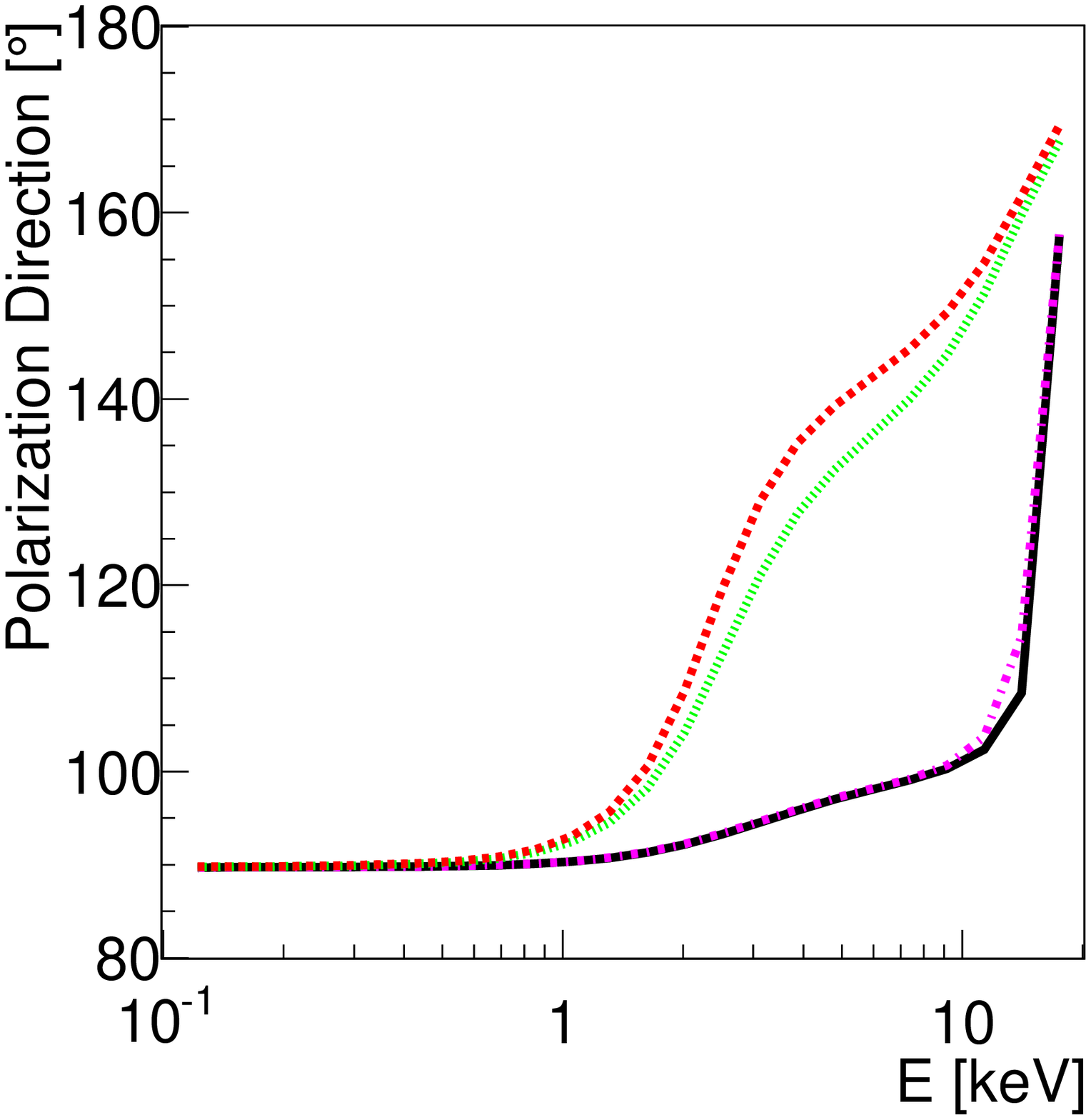}}
\end{minipage}
\caption{\label{F:C3} The same as Fig.\ \ref{F:C0} but for black holes with
$a=0$ and $\epsilon_3=0$ (solid black line, Model A),
$a=0.5$ and $\epsilon_3=-5$ (dash-dotted magenta line, Model F),
$a=0.5$ and $\epsilon_3=6.3$ (dashed red line, Model H), and
$a=0.9$ and $\epsilon_3=0$ (dotted green line, Model C).
}
\end{figure}

The GEMS mission achieves the best sensitivity ($\sim$1\% polarization degree for a mCrab source)  
with a photoelectric effect polarimeter operating over the 2-10 keV energy range \citep{Blac:10}.  
GEMS would be able to detect the polarization of the X-ray emission 
from  bright X-ray binaries like Cyg X-1 and GRS 1915$+$105 with an excellent signal to noise ratio.
For a deep ($\sim 10^6$ sec) observation of a source with a flux of 300 mCrab, statistical and systematics
errors of the polarization in 1 keV wide bins would be of the order of 0.2\% (Kallman et al., private 
communication).  The errors would be sufficiently small to allow us to distinguish between 
the Kerr models shown in Fig.\ \ref{F:C0}  \citep[see the discussion by][for more details]{Schn:10}. 
However, the GEMS sensitivity would not be sufficient to distinguish between  the GR and 
non-GR models shown in Fig.\ \ref{F:C3}. In practice, systematic errors on the model predictions 
associated with remaining uncertainties of the structure of the accretion disk and its atmosphere 
\citep[influencing e.g.\ the extent of depolarization due to Faraday rotation, and the spectral 
hardening factor $f_{\rm h}$, see][]{Davi:09}, as well as systematics associated with the subtraction of 
``contaminating'' emission components would further complicate this task.
Extending the polarimetric coverage to lower and higher energies would help to further 
constrain models. Polarimetry detectors with a broader bandpass have been
discussed in the literature \citep[e.g.][]{Weis:06,Lei:97,Bell:10,Mars:10,Kraw:11,Kraw:12}. 
Excellent sensitivity below 2 keV would make it possible to constrain the impact of 
Faraday rotation and thus the magnetic field structure in the accretion disk and its
atmosphere \citep{Davi:09}. A polarimetry mission with excellent sensitivity in the hard X-ray
band \citep[e.g. the BEST mission, see][]{Kraw:12} would allow us to characterize the polarization 
properties of the harder emission components and to reduce the errors associated with 
subtracting these components.
Other metrics may imply different observational effects, and it is clear that it would be desirable 
to explore the space of non-GR metrics in a more systematic fashion.

Testing fundamental physics with astronomical observations requires a
proper understanding of the astrophysics of the observed systems. In our case, 
we need to continue to refine accretion disk and radiation models before studies like the ones 
presented in this paper can be used to test GR. As mentioned above, numerical models are 
now used to test the more than forty year old conjecture of a vanishing torque at the ISCO,
and give more detailed information about the physical properties of accretion disks 
\citep[e.g. about the radius at which the Compton optical depth becomes smaller than unity, see][]{Krol:05}.
Even though numerical models have made impressive progress, they are still missing important 
physics including detailed modeling of the effect of radiative energy transport.   
Future studies of the predicted polarization signatures could improve on the simplified modeling 
of the emission and scattering processes adopted in this paper. Such calculations could employ a 
more detailed model of the accretion disk with information about the structure of the plasma 
and the magnetic field. More accurate modeling of the emission, scattering, and absorption 
processes could account for the composition and ionization state of the disk material, for 
different atomic transitions, multiple scattering, and Compton scattering
\citep[see][]{Nagi:62,Losk:81,Losk:85,Davi:06}. Recently, \citet{Brod:03,Brod:04}, and 
\citet{Gamm:12} developed methods for modeling the propagation of polarized rays through 
magnetized plasmas in curved spacetimes which may be used to model the polarization
of the emission from turbulent and magnetized accretion disks accounting for the 
Faraday rotation of the polarization direction.

\citet{Joha:12} studied the possibility to use Fe K-$\alpha$ fluorescent line observations 
of black holes in X-ray binaries to distinguish between the Kerr metric and the metric of Equ.\ (\ref{E:metric}).
Similarly as in this paper, they find that the parameters $a$ and $\epsilon_3$ are degenerate for some regions of the
parameter space. For these regions, distinguishing between the different metrics will require observations 
with a very high signal to noise ratio, an exquisite understanding of the accretion disk properties, 
and excellent control over systematics associated with the subtraction of the underlying continuum emission. 

\citet{Bamb:12} discusses the possibility to use the observed jet power to get an independent handle on the
black hole spin.  Assuming that black hole jets are powered by the Blandford-Znajek mechanism \citep{Blan:77}, 
the correlation between the jet power and the black hole spin can be used to break the degeneracy between
$a$ and $\epsilon_3$. Although this technique has promise, its practical utility will be limited in the foreseeable future
by our imperfect understanding of the jet launching mechanism and systematic effects, e.g.\ 
the correlation between the magnetic field close to the black hole event horizon and the black hole spin.    

The temporal variability of the observed fluxes and polarization properties offer additional diagnostics. 
For example, \citet{Hora:06a,Hora:06b} study observational signatures produced by 
moving clouds that scatter emission and emphasize that gravitational lensing can result in time delays for 
parts of the signal. It may be possible to use time resolved spectropolarimetric observations of such
transient events to constrain the geometry of the system and the underlaying spacetime.
\acknowledgments
{\it Acknowledgments:}
HK acknowledges support from NASA (grant NNX10AJ56G), and the Office of High Energy Physics 
of the US Department of Energy. The author thanks Jeremy Schnittman for participating
in detailed comparisons of results presented in this paper with results derived from his code
and for detailed comments for the manuscript.  He thanks Tim Johannsen, Julian Krolik, Jim Buckley, 
and Jonathan Katz for valuable feedback, and Clifford Will for an enjoyable GR class that the author audited.
HK acknowledges very useful comments from an anonymous referee.
\begin{deluxetable}{ccccccc}
\scriptsize
\tablecaption{Parameters describing the simulated metrics \label{T:metric}}
\tablewidth{0pt}
\tablehead{
\colhead{Model} & 
\colhead{$M\, [M_{\odot}]$}  & 
\colhead{$\dot{M}\, [10^{18} \rm \,g\, s^{-1}]$} & 
\colhead{a} & 
\colhead{$\epsilon_3$} & 
\colhead{$r_{\rm H} [r_{\rm g}]$}&
\colhead{$r_{\rm ISCO} [r_{\rm g}]$}
} \startdata
A & 10 & 2.45 & 0       & 0 & 2 & 6 \\ 
B & 10 & 1.7 & 0.5   & 0 & 1.87 & 4.23 \\
C & 10 & 0.90 & 0.9   & 0 & 1.44 & 2.32 \\
D & 10 & 0.53 & 0.99 & 0 & 1.14 & 1.45 \\  \hline
E & 10 & 4.00 & 0.5    & -30.61 & 3.08-3.13 & 10.00 \\ 
F & 10 & 2.33 & 0.5    & -5 & 1.87-1.96 & 5.79 \\ 
G & 10 & 1.27 & 0.5    & 2.5 & 1.80-1.87 & 3.28 \\ 
H & 10 & 0.88 & 0.5    & 6.33 & 1.74-1.87 & 2.32 \\  \hline  
I & 10 & 1.88 & 0.99    & -5 & 1.22-1.87 & 5.09 \\ 
J & 10 & 1.49 & 0.99    & -2.5 & 1.14-1.71 & 2.61 \\  
\hline
\enddata
\end{deluxetable}
\clearpage

\end{document}